\documentstyle[epsfig,twocolumn,subfigure]{mn}

\newcommand{\goodgap}{
\hspace{\subfigtopskip} \hspace{\subfigbottomskip}}

\title[LSB rotation curves and $R^n$ gravity theories]{Low surface brightness galaxies rotation curves in the
low energy limit of $R^n$ gravity\,: no need for dark matter ?}
\author[S. Capozziello et al.]{S. Capozziello$^{1}$, V.F. Cardone$^2$, A. Troisi$^1$ \\
$^1$ Dipartimento di Scienze Fisiche, Universit\`{a} degli studi
di Napoli ``Federico II'' and INFN, Sezione di Napoli, \\
Complesso Universitario di Monte S. Angelo, Via Cinthia, Edificio
N, 80126 Napoli, Italy\\ $^2$ Dipartimento di Fisica ``E.R.
Caianiello'', Universit{\`{a}} di Salerno, \\ Via S. Allende,
84081 - Baronissi (Salerno), Italy}

\date{Accepted xxx, Received yyy, in original form zzz}

\begin{document}

\maketitle

\begin{abstract}

We investigate the possibility that the observed flatness of the
rotation curves of spiral galaxies is not an evidence for the
existence of dark matter haloes, but rather a signal of the
breakdown of General Relativity. To this aim, we consider
power\,-\,law fourth order theories of gravity obtained by
replacing the scalar curvature $R$ with $f(R) = f_0 R^n$ in the
gravity Lagrangian. We show that, in the low energy limit, the
gravitational potential generated by a pointlike source may be
written as $\Phi(r) \propto r^{-1} \left [ 1 + (r/r_c)^{\beta}
\right ]$ with $\beta$ a function of the slope $n$ of the gravity
Lagrangian and $r_c$ a scalelength depending on the gravitating
system properties. In order to apply the model to realistic
systems, we compute the modified potential and the rotation curve
for spherically symmetric and for thin disk mass distributions. It
turns out that the potential is still asymptotically decreasing,
but the corrected rotation curve, although not flat, is higher
than the Newtonian one thus offering the possibility to fit
rotation curves without dark matter. To test the viability of the
model, we consider a sample of 15 low surface brightness (LSB)
galaxies with combined HI and H$\alpha$ measurements of the
rotation curve extending in the putative dark matter dominated
region. We find a very good agreement between the theoretical
rotation curve and the data using only stellar disk and
interstellar gas when the slope $n$ of the gravity Lagrangian is
set to the value $n = 3.5$ (giving $\beta = 0.817$) obtained by
fitting the SNeIa Hubble diagram with the assumed power\,-\,law
$f(R)$ model and no dark matter. The excellent agreement among
theoretical and observed rotation curves and the values of the
stellar mass\,-\,to\,-\,light ratios in agreement with the
predictions of population synthesis models make us confident that
$R^n$ gravity may represent a good candidate to solve both the
dark energy problem on cosmological scales and the dark matter one
on galactic scales with the same value of the slope $n$ of the
higher order gravity Lagrangian.

\end{abstract}


\section{Introduction}

An impressive amount of unprecedented high quality data have been
accumulated in the last decade and have depicted the new picture
of a spatially flat universe with a subcritical matter content and
undergoing a phase of accelerated expansion. The measurements of
cluster properties as the mass and correlation function and the
evolution with redshift of their abundance
\cite{eke98,vnl02,bach03,bb03}, the Hubble diagram of Type Ia
Supernovae \cite{Riess04,ast05,clo05}, optical surveys of large
scale structure \cite{pope04,cole05,eis05}, anisotropies in the
cosmic microwave background \cite{Boom,WMAP}, cosmic shear from
weak lensing surveys \cite{vW01,refr03} and the Lyman\,-\,$\alpha$
forest absorption \cite{chd99,mcd04} are concordant evidences in
favour of the radically new scenario depicted above. Interpreting
this huge (and ever increasing) amount of information in the
framework of a single satisfactory theoretical model is the main
challenge of modern cosmology.

Although it provides an excellent fit to the most of the data
\cite{Teg03,Sel04,sanch05}, the old cosmological constant
\cite{CarLam,Sahni} is affected by serious theoretical
shortcomings that have motivated the search for alternative
candidates generically referred to as {\it dark energy}. Rather
than enumerating the many ideas on the ground (from a scalar field
rolling down a suitably chosen self interaction potential to
phantom fields and unified models of dark energy and dark matter),
we refer the interested reader to the enlightening reviews
available in literature (see, e.g., Peebles \& Rathra 2003 and
Padmanabhan 2003). Here, we only remind that dark energy acts as a
negative pressure fluid whose nature and fundamental properties
remain essentially unknown notwithstanding the great theoretical
efforts made up to now.

Rather than being evidence for the need of some unknown component
in the energy budget, the cosmic speed up of a low matter universe
may also be considered as a first signal of breakdown of Einstein
General Relativity. In this framework, higher order theories of
gravity represent an interesting opportunity to explain cosmic
acceleration without the need of any dark energy. In such models,
the Ricci scalar curvature $R$ in the gravity Lagrangian is
replaced by a generic function $f(R)$ thus leading to modified
Friedmann equations that can be formally written in the usual form
by defining an effective negative pressure {\it curvature fluid}
driving the cosmic acceleration
\cite{capozzcurv,noirev,cdtt,no03a}. Also referred to as $f(R)$
theories, this approach has been extensively studied both from the
theoretical (see, e.g., Capozziello, Cardone and Troisi 2005 and
refs. therein) and observational point of view
\cite{noiijmpd,ccf,borowiec}. Moreover, this same approach has
been also proposed as a mechanism to give rise to an inflationary
era without the need of any inflaton field \cite{Star80}. All
these works have been concentrated on the cosmological
applications of $f(R)$ theories and have convincingly demonstrated
that they are indeed able to explain the cosmic speed up and fit
the available dataset and hence represents a viable alternative to
the mysterious dark energy.

Changing the gravity Lagrangian has consequences not only on
cosmological scales, but also at the galactic ones so that it is
mandatory to investigate the low energy limit of $f(R)$ theories.
Unfortunately, here a strong debate is still open with different
papers drawing contrasting results arguing in favour
\cite{dick,sotiriou,cembranos,navarro,allrugg,ppnantro} or against
\cite{dolgov,chiba,olmo} such models. It is worth noting that, as
a general result, higher order theories of gravity cause the
gravitational potential to deviate from its Newtonian $1/r$
scaling \cite{stelle78,hj,hjrev,cb05,sobouti} even if such
deviations may also be very soon vanishing.

In a previous paper \cite{noipla},  the Newtonian limit of power
law $f(R) = f_0 R^n$ theories has been investigated, assuming that
the metric in the low energy limit ($\Phi/c^2 << 1$) may be taken
as Schwarzschild\,-\,like. It turns out that a power law term
$(r/r_c)^{\beta}$ has to be added to the Newtonian $1/r$ term in
order to get the correct gravitational potential. While the
parameter $\beta$ may be expressed analytically as function of the
slope $n$ of the $f(R)$ theory, $r_c$ sets the scale where the
correction term starts being significant and has to be determined
case\,-\,by\,-\,case. We then investigated a particular range of
values of $n$ leading to $\beta > 0$ so that the corrective term
is an increasing function of the radius $r$ thus causing an
increase of the rotation curve with respect to the Newtonian one
and offering the possibility to fit the galaxy rotation curves
without the need of the elusive dark matter component. As a
preliminary test, we successfully fitted the Milky Way rotation
curve using a model made out of the luminous components (bulge and
disk) only. Notwithstanding these encouraging results, the
corrected potential for $n$ in the range explored in our previous
paper is, however, troublesome. Indeed, the correction term never
switches off so that the total gravitational potential has the
unpleasant feature of being formally divergent as $r$ goes to
infinity. Actually, the expression for the gravitational potential
has been obtained in the low energy limit so that cannot be
extrapolated to distances where this approximation does not hold
anymore. Nevertheless, for typical values of the parameters
$(\beta, r_c)$, the rotation curve starts increasing for values of
$r$ near the visible edge of the disk thus contradicting what is
observed in outer galaxies where the rotation curve is flat or
slowly rising \cite{pss96,catinella}.

Elaborating further on the previous results, we present here the
analysis of the gravitational potential which is obtained by
considering a different approach to the Newtonian limit of $f(R)$
theories giving rise to a correction term $(r/r_c)^{\gamma}$ with
$-1 < \gamma < 0$. As we will see, the corresponding rotation
curve is asymptotically decreasing as in the Newtonian case, but
is nevertheless higher than the standard one so that it is still
possible to fit the data without the need of dark matter.
Moreover, for such models, the gravitational potential
asymptotically vanishes so that the problem discussed above is
avoided. To further substantiate our model, we consider a set of
low surface brightness (hereafter LSB) galaxies with extended and
well measured rotation curves. Since these systems are supposed to
be dark matter dominated, successfully fitting our model with {\it
no dark matter} to the LSB rotation curves would be a strong
evidence in favor of our approach. Combined with the hints coming
from the cosmological applications discussed above, we should thus
have the possibility to solve both the dark energy and dark matter
problems resorting to the same well motivated fundamental theory
(see \cite{prl} for preliminary results in this sense).

The plan of the paper is as follows. In Sect.\,2, we briefly
resume how the gravitational potential may be obtained in the low
energy limit of power\,-\,law $f(R)$ theories in the case of a
pointlike source. The generalization to both a spherically
symmetric system and a thin disk is presented in Sect.\,3. The
data on the rotation curve, the modelling of LSB galaxies and the
method adopted to determine the model parameters are presented in
Sect.\,4. An extensive analysis of the fitting procedure is
carried out in Sect.\,5 where we use simulated rotation curves to
investigate how parameter degeneracies affect the estimate of the
model parameters. The results of the fit are presented in
Sect.\,6, while Sect.\,7 is devoted to summarize and foresee
future prospects. Some more details on the fit results on a
case\,-\,by\,-\,case basis and on the smoothing procedure are
given in Appendix A and B respectively.

\section{Low energy limit of $f(R)$ gravity}

As yet stated in the introduction, $f(R)$ theories of gravity
represent a straightforward generalization of the Einstein General
Relativity. To this aim, one considers the action\,:

\begin{equation}
\label{f(R)action} {\cal{A}} = \int{d^4x \sqrt{-g} \left [ f(R) +
{\cal{L}}_{m} \right ]}
\end{equation}
where $f(R)$ is a generic analytic function of the Ricci scalar
curvature $R$ and ${\cal{L}}_m$ is the standard matter Lagrangian.
The choice $f(R) = R + 2\Lambda$ gives the General Relativity
including the contribution of the cosmological constant $\Lambda$.
Varying the action with respect to the metric components $g_{\mu
\nu}$, one gets the generalized Einstein equations that can be
more expressively recast as \cite{capozzcurv,cct}\,:

\begin{eqnarray}
G_{\mu \nu} & = & \displaystyle{\frac{1}{f'(R)}}
\displaystyle{\Bigg \{ \frac{1}{2} g_{\mu \nu} \left [ f(R) - R
f'(R) \right ] + f'(R)_{; \mu \nu}} \nonumber \\ ~ & - &
\displaystyle{g_{\mu \nu} \Box{f'(R)} \Bigg \}} +
\displaystyle{\frac{T^{(m)}_{\mu \nu}}{f'(R)}} \label{eq:f-var2}
\end{eqnarray}
where $G_{\mu\nu} = R_{\mu \nu} - (R/2) g_{\mu \nu}$ is the
Einstein tensor and the prime denotes derivative with respect to
$R$. The two terms ${f'(R)}_{; \mu \nu}$ and $\Box{f'(R)}$ imply
fourth order derivatives of the metric $g_{\mu \nu}$ so that these
models are also referred to as {\it fourth order gravity}.
Starting from Eq.(\ref{eq:f-var2}) and adopting the
Robertson\,-\,Walker metric, it is possible to show that the
Friedmann equations may still be written in the usual form
provided that an {\it effective curvature fluid} (hence the name
of {\it curvature quintessence}) is added to the matter term with
energy density and pressure depending on the choice of $f(R)$. As
a particular case, we consider power\,-\,law $f(R)$ theories, i.e.
we set\,:

\begin{equation}
f(R) = f_0 R^n \label{eq: frn}
\end{equation}
with $n$ the slope of the gravity Lagrangian ($n = 1$ being the
Einstein theory) and $f_0$ a constant with the dimensions chosen
in such a way to give $f(R)$ the right physical dimensions. It has
been shown that the choice (\ref{eq: frn}), with $n \neq 1$ and
standard matter, is able to properly fit the Hubble diagram of
Type Ia Supernovae without the need of dark energy
\cite{noiijmpd,sante} and could also be reconciled with the
constraints on the PPN parameters \cite{ppnantro}.

Here we study the low energy limit\footnote{Although not
rigourously correct, in the following we will use the terms {\it
low energy limit} and {\it Newtonian limit} as synonymous.} of
this class of $f(R)$ theories. Let us consider the gravitational
field generated by a pointlike source and solve the field
equations (\ref{eq:f-var2}) in the vacuum case. Under the
hypothesis of weak gravitational fields and slow motions, we can
write the spacetime metric as\,:

\begin{equation}
ds^2 = A(r) dt^2 - B(r) dr^2 - r^2 d\Omega^2 \label{eq: schwartz}
\end{equation}
where $d\Omega^2 = d\theta^2 + \sin^2{\theta} d\varphi^2$ is the
line element on the unit sphere. It is worth noting that writing
Eq.(\ref{eq: schwartz}) for the weak field metric is the same as
assuming implicitly that the Jebsen\,-\,Birkhoff theorem holds.
While this is true in standard General Relativity, it has never
been definitively proved for $f(R)$ theories. Actually, since for
a general $f(R)$ theory the field equations are fourth order, it
is quite difficult to show that the only stationary spherically
symmetric vacuum solution is Schwarzschild like. However, that
this is indeed the case has been demonstrated for $f(R)$ theories
involving terms like $R + R^2$ with $R^2 = R^{\alpha}_{\ \beta}
{^{\mu \nu}} R_{\alpha}^{\ \beta} {_{\mu \nu}}$ with torsion
\cite{yass} and for the case of any invariant of the form $R^2$
also in the case of null torsion \cite{nev}. Moreover, the
Jebsen\,-\,Birkhoff theorem has been shown to hold also for more
complicated theories as multidimensional gravity and
Einstein\,-\,Yang\,-\,Mills theories \cite{bs93,bm94}. Therefore,
although a rigorous demonstration is still absent, it is likely
that this theorem is still valid for power\,-\,law $f(R)$
theories, at least in an {\it approximated} weak
version\footnote{It is, for instance, possible that the metric
(\ref{eq: schwartz}) solves the field equations only up to terms
of low order in $\Phi/c^2$ with $\Phi$ the gravitational
potential. For the applications we are interested in, $\Phi/c^2 <<
1$, such weak version of the Jebsen\,-\,Birkhoff theorem should be
verified.} that is enough for our aims.

To find the two unknown functions $A(r)$ and $B(r)$, we first
combine the $00$\,-\,vacuum component and the trace of the field
equations (\ref{eq:f-var2}) in absence of matter\,:

\begin{displaymath}
3 \Box{f'(R)} + R f'(R) - 2 f(R) = 0 \ ,
\end{displaymath}
to get a single equation\,:

\begin{equation}
\label{master-low} f'(R) \left ( 3 \frac{R_{00}}{g_{00}} - R
\right ) + \frac{1}{2} f(R) - 3 \frac{f'(R)_{; 0 0}}{g_{00}} = 0 \
.
\end{equation}
Eq.(\ref{master-low}) is completely general and holds whatever is
$f(R)$. It is worth stressing, in particular, that, even if the
metric is stationary so that $\partial_{t} g_{\mu \nu} = 0$, the
term $f'(R)_{; 0 0}$ is not vanishing because of the non-null
Christoffel symbols entering the covariant derivative. Using
Eq.(\ref{eq: frn}), Eq.(\ref{master-low}) reduces to\,:

\begin{equation}
\label{master-pla} R_{00}(r) = \frac{2 n - 1}{6 n} \ A(r) R(r) -
\frac{n - 1}{2 B(r)} \frac{dA(r)}{dr} \frac{d\ln{R(r)}}{dr} \ ,
\end{equation}
while the trace equation reads\,:

\begin{equation}
\Box{R^{n - 1}(r)} = \frac{2 - n}{3 n} R^n(r) \ . \label{eq:
tracebis}
\end{equation}
Note that for $n = 1$, Eq.(\ref{eq: tracebis}) reduces to $R = 0$,
which, inserted into Eq.(\ref{master-pla}), gives $R_{00} = 0$ and
the standard Schwarzschild solution is recovered. In general,
expressing $R_{00}$ and $R$ in terms of the metric (\ref{eq:
schwartz}), Eqs.(\ref{master-pla}) and (\ref{eq: tracebis}) become
a system of two nonlinear coupled differential equations for the
two functions $A(r)$ and $B(r)$. A physically motivated hypothesis
to search for solutions is

\begin{equation}
A(r) = \frac{1}{B(r)} = 1 + \frac{2 \Phi(r)}{c^2} \label{eq:
avsphi}
\end{equation}
with $\Phi(r)$ the gravitational potential generated by a
pointlike mass $m$ at the distance $r$. With the above hypothesis,
the vacuum field equations reduce to a system of two differential
equations in the only unknown function $\Phi(r)$. To be more
precise, we can solve Eq.(\ref{master-pla}) or (\ref{eq:
tracebis}) to find out $\Phi(r)$ and then use the other relation
as a constraint to find  solutions of physical interest. To this
aim, let us remember that, as well known, $f(R)$ theories induces
modifications to the gravitational potential altering the
Newtonian $1/r$ scaling \cite{stelle78,hj,hjrev,cb05}. We thus
look for a solution for the potential that may be written as\,:

\begin{equation}
\Phi(r) = - \frac{G m}{2 r} \left [ 1 + \left ( \frac{r}{r_c}
\right )^{\beta} \right ] \label{eq: pointphi}
\end{equation}
so that the gravitational potential deviates from the usual
Newtonian one because of the presence of the second term on the
right hand side. Note that, when $\beta = 0$, the Newtonian
potential is recovered and the metric reduces to the classical
Schwarzschild one. On the other hand, as we will see, it is just
this term that offers the intriguing possibility to fit galaxy
rotation curves without the need of dark matter.

In order to check whether Eq.(\ref{eq: pointphi}) is indeed a
viable solution, we first insert the expression for $\Phi(r)$ into
Eqs.(\ref{master-pla}) and (\ref{eq: tracebis}) which are both
solved if\,:

\begin{eqnarray}
&&(n - 1) (\beta - 3) \left [- \beta (1 + \beta) V_1 \eta^{\beta -
3} \right ]^{n - 1}\nonumber\\ &&{\times}\left [1+ \frac{\beta V_1
{\cal{P}}_0}{{\cal{P}}_1\eta} \right]{\cal{P}}_1\eta = 0
\label{eq: singleeq}
\end{eqnarray}
with $\eta = r/r_c$, $V_1 = G m/ c^2 r_c$ and

\begin{eqnarray}
{\cal{P}}_0 & = & 3 (\beta - 3)^2 n^3  - (5 \beta^2 - 31
\beta + 48) n^2 \nonumber \\
~ & - & (3 \beta^2 - 16 \beta + 17) n - (\beta^2 - 4 \beta - 5) \
, \label{eq: pz}
\end{eqnarray}

\begin{eqnarray}
{\cal{P}}_1 & = & 3 (\beta - 3)^2 (1 - \beta) n^3  +
(\beta - 3)^2 (5 \beta - 7) n^2 \nonumber \\
~ & - & (3 \beta^3 - 17 \beta^2 + 34 \beta - 36) n + (\beta^2 - 3
\beta - 4) \beta \ . \label{eq: pu}
\end{eqnarray}
Eq.(\ref{eq: singleeq}) is identically satisfied for  particular
values of $n$ and $\beta$. However, there are some simple
considerations allow to exclude such values. First, $n = 1$ must
be discarded since, when deriving Eq.(\ref{master-pla}) from
Eq.(\ref{master-low}), we have assumed $R \ne 0$ which is not the
case for $n = 1$. Second, the case $\beta = 3$ may also be
rejected since it gives rise to a correction to the Newtonian
potential scaling as $\eta^2$ so that the total potential diverges
quadratically which is quite problematic. Finally, the case $\beta
= -1$ provides a solution only if $n > 1$. Since we are interested
in a solution which works whatever $n$ is, we discard also this
case. However, in the limit we are considering, it is $V_1 << 1$.
For instance, it is $V_1 \simeq v_c^2/c^2 \sim 10^{-6}\div
10^{-3}$ ranging from Solar System to galactic scales, with $v_c$
the circular velocity. As a consequence, we can look for a further
solution of Eq.(\ref{eq: singleeq}) solving\,:

\begin{equation}
{\cal{P}}_1(n, \beta) \eta + \beta V_1 {\cal{P}}_0(n, \beta)
\simeq {\cal{P}}_1(n, \beta) \eta = 0 \label{eq: singlepol} \ .
\end{equation}
since the second term of  Eq.(\ref{eq: singlepol}) is always
negligible for the values of $n$ and $\beta$ in which we are
interested.  Eq.(\ref{eq: singlepol}) is an algebraic equation for
$\beta$ as function of $n$ with the following three solutions\,:

\begin{equation}
\beta = \left \{
\begin{array}{l}
\displaystyle{\frac{3n - 4}{n - 1}} \\
\displaystyle{\frac{12n^2 - 7n - 1 - \sqrt{p(n)}}{q(n)}} \\
\displaystyle{\frac{12n^2 - 7n - 1 + \sqrt{p(n)}}{q(n)}} \\
\end{array} \right .
\label{eq: bn}
\end{equation}
with\,:

\begin{displaymath}
p(n) = 36n^4 + 12n^3 - 83n^2 + 50n + 1 \ ,
\end{displaymath}

\begin{displaymath}
q(n) = 6n^2 - 4n + 2 \ .
\end{displaymath}
It is easy check that, for $n = 1$,  the second expression gives
$\beta = 0$, i.e. the approximate solution reduces to the
Newtonian one as expected. As a final check, we have inserted back
into the vacuum field equations (\ref{master-low}) and (\ref{eq:
tracebis}) the modified gravitational potential (\ref{eq:
pointphi}) with

\begin{equation}
\beta = \frac{12n^2 - 7n - 1 - \sqrt{36n^4 + 12n^3 - 83n^2 + 50n +
1}}{6n^2 - 4n + 2} \label{eq: bnfinal}
\end{equation}
finding out that the approximated solution solve the field
equations up to $10^{-6}$ which is more than sufficient in all
astrophysical applications which we are going to consider.

Armed with Eqs.(\ref{eq: pointphi}) and (\ref{eq: bnfinal}), we
can, in principle, set constraints on $n$ by imposing some
physically motivated requirements to the modified gravitational
potential. However, given the nonlinear relation between $n$ and
$\beta$, in the following we will consider $\beta$ and use
Eq.(\ref{eq: bnfinal}) to infer $n$ from the estimated $\beta$.

As a first condition, it is reasonable to ask that the potential
does not diverge at infinity. To this aim, we impose\,:

\begin{displaymath}
\lim_{r \rightarrow \infty}{\Phi(r)} = 0
\end{displaymath}
which constraints $\beta - 1$ to be negative. A further constraint
can be obtained considering the Newtonian potential $1/r$ as valid
at Solar System scales. As a consequence, since the correction to
the potential scales as $r^{\beta - 1}$, we must impose $\beta - 1
> -1$ in order to avoid increasing $\Phi$ at the Solar System
scales. In order to not evade these constraints, in the following,
we will only consider solutions with

\begin{equation}
0 < \beta < 1 \  \label{eq: brange}
\end{equation}
that, using Eq.(\ref{eq: bnfinal}), gives $n > 1$ as lower limit
on the slope $n$ of the gravity Lagrangian.

While $\beta$ controls the shape of the correction term, the
parameter $r_c$ controls the scale where deviations from the
Newtonian potential sets in. Both $\beta$ and $r_c$ have to be
determined by comparison with observations at galactic scales. An
important remark is in order here. Because of Eq.(\ref{eq:
bnfinal}), $\beta$ is related to $n$ which enters the gravity
Lagrangian. Since this is the same for all gravitating systems, as
a consequence, $\beta$ must be the same for all galaxies. On the
other hand, the scalelength parameter $r_c$ is related to the
boundary conditions and the mass of the system. In fact,
considering the generalization of Eq.(\ref{eq: pointphi}) to
extended systems, one has to take care of the mass distribution
and the geometrical configurations which can differ from one
galaxy to another. As a consequence, $r_c$ turns out to be not a
universal quantity, but its value must be set on a
case\,-\,by\,-\,case basis.

Before considering the generalization to extended systems, it is
worth evaluating the rotation curve for the pointlike case, i.e.
the circular velocity $v_c(r)$ of a test particle in the potential
generated by the point mass $m$. For a central potential, it is
$v_c^2 = r d\Phi/dr$ that, with $\Phi$ given by Eq.(\ref{eq:
pointphi}), gives\,:

\begin{equation}
v_c^2(r) = \frac{G m}{2 r} \left [ 1 + (1 - \beta) \left (
\frac{r}{r_c} \right )^{\beta} \right ] \ . \label{eq: vcpoint}
\end{equation}
As it is apparent, the corrected rotation curve is the sum of two
terms. While the first one equals half the Newtonian curve $G m
/r$, the second one gives a contribution that may also be higher
than the half classical one. As expected, for $\beta = 0$, the two
terms sum up to reproduce the classical result. On the other hand,
for $\beta$ in the range (\ref{eq: brange}), $1 - \beta
> 0$ so that the the corrected rotation curve is higher than the
Newtonian one. Since measurements of spiral galaxies rotation
curves signal a circular velocity higher than what is predicted on
the basis of the observed mass and the Newtonian potential, the
result above suggests the possibility that our modified
gravitational potential may fill the gap between theory and
observations without the need of additional dark matter.

It is worth noting that the corrected rotation curve is
asymptotically vanishing as in the Newtonian case, while it is
usually claimed that observed rotation curves are flat (i.e.,
asymptotically constant). However, such a statement should be not
be taken literally. Actually, observations do not probe $v_c$ up
to infinity, but only up to a maximum radius $R_{max}$ showing
that the rotation curve is flat within the measurement
uncertainties. However, this by no way excludes the possibility
that $v_c$ goes to zero at infinity. Considering Eq.(\ref{eq:
vcpoint}), if the exponent of the correction term is quite small,
the first term decreases in a Keplerian way, while the second one
approaches its asymptotically null value very slowly so that it
can easily mimic an approximately flat rotation curve in agreement
with observations.

\section{Extended systems}

The solution (\ref{eq: pointphi}) has been obtained in the case of
a pointlike source, but may be easily generalized to the case of
extended systems. To this aim, we may simply divide the system in
infinitesimal elements with mass $dm$ and add the different
contributions. In the continuous limit, the sum is replaced by an
integral depending on the mass density and the symmetry of the
system spatial configuration. Once the gravitational potential has
been obtained, the rotation curve may be easily evaluated and then
compared with observations.

\subsection{Spherically symmetric systems}

The generalization of Eq.(\ref{eq: pointphi}) to a spherically
symmetric system is less trivial than one would expect. In the
case of the Newtonian gravitational potential, the Gauss theorem
ensures us that the flux of the gravitational field generated by a
point mass $m$ through a closed surface only depends on the mass
$m$ and not on the position of the mass inside the surface.
Moreover, the force on a point inside the surface due to sources
outside the surface vanishes. As a result, we may imagine that the
whole mass of the system is concentrated in its centre and, as a
consequence, the gravitational potential has the same formal
expression as for the pointlike case provided one replaces $m$
with $M(r)$, being this latter quantity the mass within a distance
$r$ from the centre.

From a mathematical point of view, we can write in the Newtonian
case \,:

\begin{eqnarray}
\Phi_N(r) & = & - G \int{\frac{\rho({\bf x}')}{|{\bf x} - {\bf x}'|} d^3x'} \nonumber \\
~ & = & - \frac{4 \pi G}{r} \int_0^\infty{\rho(r') r'^2 dr'}
\nonumber \\ ~ & = & - \frac{G M(r)}{r}  \nonumber
\end{eqnarray}
where, in the second row, we have  used the Gauss theorem to take
the $|{\bf x} - {\bf x}'|^{-1}$ outside the integral sign
(considering all the mass  concentrated in the point ${\bf x}' =
0$) and then limited  the integral to $r$ since points with $r'
> r$ do not contribute to the gravitational force.

It is quite easy to show that the Gauss theorem for the
gravitational field is a consequence of the scaling $1/r^2$ of the
Newtonian force. Since this scaling is lost in the case of the
modified potential (\ref{eq: pointphi}), the Gauss theorem does
not hold anymore. However, apart from the multiplicative factor
$1/2$, we can split the modified gravitational potential as the
sum of two terms, the first one scaling as in the Newtonian case.
For this term, the Gauss theorem holds and we recover the
classical results so that the total gravitational potential of a
spherically symmetric system may be written as\,:

\begin{equation}
\Phi(r) =  \frac{\Phi_N(r) + \Phi_c(r)}{2} = - \frac{G M(r)}{2 r}
+ \frac{\Phi_c(r)}{2} \label{eq: phispher}
\end{equation}
with\,:

\begin{equation}
\Phi_c(r) = -G \int_{0}^{\infty}{\rho(r') r'^2 dr'}
\int_{-\pi/2}^{\pi/2}{\sin{\theta'} d\theta'} \int_{0}^{2
\pi}{\psi_c d\phi'} \label{eq: psiccap}
\end{equation}
with $\psi_c$ the non-Newtonian part of the modified gravitational
potential for the pointlike case. In order to be  more general, we
consider the calculation for a generic modified potential of the
type\,:

\begin{equation}
\psi = - \frac{G m}{2 r} \left [1 + \alpha \left ( \frac{r}{r_c}
\right )^{\beta} \right ] \label{eq: genphipoint}
\end{equation}
so that\,:

\begin{equation}
\psi_c(r) = - \frac{\alpha G m}{r_c} \left ( \frac{r}{r_c} \right
)^{\beta - 1}
\label{eq: psicdef}
\end{equation}
with $\alpha$ and $\beta$ two parameters depending on the
particular theory of gravity one is considering. While for $R^n$
gravity $\alpha = 1$, in general, $\alpha$ could also be negative.
Inserting the above $\psi_c$ into Eq.(\ref{eq: psiccap}), we
replace $r'$ with

\begin{displaymath}
|{\bf x} - {\bf x}'| = (r^2 + r'^2 - 2 r r' \cos{\theta'})^{1/2}
\end{displaymath}
where we have used the spherical symmetry of the system so that
the potential in the point ${\bf x} = (r, \theta, \phi)$ only
depends on $r$ and we can set $\theta = \phi = 0$. Integrating
over the angular variables $(\theta', \phi')$, we finally get\,:

\begin{equation}
\Phi_c(r) = - \frac{\pi G \alpha r_c^2}{3} \left [ {\cal{I}}_1(r)
+ {\cal{I}}_2(r) \right ] \label{eq: phicspher}
\end{equation}
with\,:

\begin{eqnarray}
{\cal{I}}_1 & = & 3 \pi \int_{0}^{\infty}{(\xi^2 + \xi'^2)^{(\beta
- 1)/2}
\rho(\xi') \xi'^2 d\xi'} \nonumber \\
~ & {\times} & {_2F_1}\left [\left \{ \frac{1 - \beta}{4}, \frac{3
- \beta}{4} \right\}, \{2\}, \frac{4 \xi^2 \xi'^2}{(\xi^2 +
\xi'^2)^2} \right ] \ , \label{eq: defione}
\end{eqnarray}

\begin{eqnarray}
{\cal{I}}_2 & = & 4 (1 - \beta) \xi \int_{0}^{\infty}{ (\xi^2 +
\xi'^2)^{(\beta - 3)/2} \rho(\xi') \xi'^2 d\xi'} \nonumber \\
~ & {\times} & {_3F_2}\left [\left \{ 1, \frac{3 - \beta}{4},
\frac{5 - \beta}{4} \right \}, \left \{ \frac{3}{2}, \frac{5}{2}
\right \}, \frac{4 \xi^2 \xi'^2}{(\xi^2 + \xi'^2)^2} \right ]  \ ,
\label{eq: defitwo}
\end{eqnarray}
and we have generically defined $\xi = r/r_c$ and used the
notation  ${_pF_1}[\{a_1, \ldots, a_p\}, \{b_1, \ldots, b_q\}, x]$
for the hypergeometric functions.

Eqs.(\ref{eq: defione}) and (\ref{eq: defitwo}) must be evaluated
numerically for a given expression of the mass density $\rho(r)$.
Once $\Phi_c(r)$ has been evaluated, we can compute the rotation
curve as\,:

\begin{equation}
v_c^2(r) = r \frac{\partial \Phi}{\partial r} =
\frac{v_{c,N}^2(r)}{2} + \frac{r}{2} \frac{\partial
\Phi_c}{\partial r}
\end{equation}
with $v_{c,N}^2(r) = G M(r)/r$ the Newtonian rotation curve. Since
we are mainly interested in spiral galaxies without any spherical
component, we do not evaluate the rotation curve explicitly. We
only note that, since $\Phi_c$ has to be evaluated numerically, in
order to avoid numerical derivatives, it is better to first
differentiate analytically the expressions for ${\cal{I}}_1$ and
${\cal{I}}_2$ and then integrate numerically the corresponding
integrals. It is easy to check that the resulting rotation curve
is typically slowly decreasing so that it vanishes asymptotically
as in the Newtonian case. However, the rate of decline is slower
than the Keplerian one so that the total rotation curve turns out
to be higher than the Newtonian one: this fact allows to fit
galaxy rotation curves without the need of any dark matter
halo\footnote{It is worth stressing, at this point, that general
conservation laws are guaranteed by Bianchi identities which hold
for generic $f(R)$, so the non-validity of Gauss theorem is not a
shortcoming since we are  considering the low energy limit of the
theory.}.

\subsection{Thin disk}

The case of a disk\,-\,like system is quite similar to the
previous one and, indeed, the gravitational potential may be
determined following the same method as before simply taking care
of the cylindrical rather than spherical symmetry of the mass
configuration. In order to simplify computations, but still
dealing with realistic systems, we will consider a circularly
symmetric and infinitesimally thin disk and denote by $\Sigma(R)$
its surface mass density\footnote{Here, $R$ is the cylindrical
coordinate in the plane of the disk (i.e., $R^2 = x^2 + y^2$) to
be not confused with the Ricci scalar curvature.} and by $R_d$ its
scale length. Note that a thin circular disk is the standard
choice in describing spiral galaxies so that the model we consider
is indeed the most realistic one.

Adopting cylindrical coordinates $(R, \phi, z)$, the gravitational
potential may be evaluated as\,:

\begin{equation}
\Phi(R, z) = \int_{0}^{\infty}{\Sigma(R') R' dR' \int_{0}^{2
\pi}{\psi(|{\bf x} - {\bf x}'|) d\phi'}} \label{eq: diskgen}
\end{equation}
with $\psi$ the pointlike potential and\,:

\begin{equation}
|{\bf x} - {\bf x}'|^2 = \left [ (R + R')^2 + z^2 \right ] \left [
1 - k^2 \cos^2{(\phi'/2)} \right ] \ , \label{eq: coord}
\end{equation}

\begin{equation}
k^2 \equiv \frac{4 R R'}{\left [ (R + R')^2 + z^2 \right ]} \ .
\label{eq: defk}
\end{equation}
Inserting Eq.(\ref{eq: genphipoint}) into Eq.(\ref{eq: diskgen}),
we get an integral that can be split into two additive terms. The
first one is half the usual Newtonian one that can be solved using
standard procedure \cite{BT87} and therefore will not be
considered anymore. The second one is the correction term
$\Phi_{c}$ that reads\footnote{As in the previous paragraph, it is
convenient to let apart the multiplicative factor $1/2$ and inser
it only in the final result so that the total potential reads
$\Phi(R, z) = \left [ \Phi_N(R, z) + \Phi_c(R, z) \right ]/2$.}\,:

\begin{eqnarray}
\Phi_c(R, z) & = &  -\alpha G \Sigma_0 r_c
\int_{0}^{\infty}{\tilde{\Sigma}(\xi') \left [ (\xi + \xi')^2 +
\zeta^2 \right ]^{\frac{\beta - 1}{2}} \xi' d\xi'} \nonumber \\ ~
& {\times} & \int_{0}^{2 \pi}{\left [ 1 - k^2 \cos^2{(\phi'/2)}
\right ]^{\frac{\beta - 1}{2}} d\phi'}
\end{eqnarray}
with $\Sigma_0 = \Sigma(R = 0)$, $\tilde{\Sigma} =
\Sigma/\Sigma_0$, $\xi = R/r_c$ and $\zeta = z/r_c$. Integrating
over $d\phi'$ and using Eq.(\ref{eq: defk}), we finally get\,:

\begin{eqnarray}
\Phi_c(R, z) & = & - 2^{\beta - 2} \pi \alpha G \Sigma_0 r_c
\xi^{\frac{\beta - 1}{2}} \int_{0}^{\infty}{d\xi' \tilde{\Sigma}(\xi') \xi'^{\frac{1 + \beta}{2}}} \nonumber \\
~ & {\times} & {_2F_1}\left [\left \{ \frac{1}{2}, \frac{1 -
\beta}{2} \right\}, \{1\}, k^2 \right ] k^{1 - \beta} \ .
\label{eq: phidiskcorr}
\end{eqnarray}
Eq.(\ref{eq: phidiskcorr}) makes it possible to evaluate the
corrective term to the gravitational potential generated by an
infinitely thin disk given its surface density $\Sigma(\xi)$. As a
useful application, we consider the case of the exponential disk
\cite{Freeman70}\,:

\begin{equation}
\Sigma(R) = \Sigma_0 \exp{\left ( - R/R_d \right )} \label{eq:
sigexp}
\end{equation}
with $R_d$ the scale radius. With this expression for the surface
density, the corrective term in the gravitational potential may be
conveniently written as\,:

\begin{eqnarray}
\Phi_c(R, z) & = & - 2^{\beta - 2} \eta_c^{-\beta} \pi \alpha G
\Sigma_0 R_d \eta^{\frac{\beta - 1}{2}} \int_{0}^{\infty}{d\eta'
{\rm e}^{-\eta'} \ \eta'^{\frac{\beta + 1}{2}}} \nonumber
\\
~ & {\times} & {_2F_1}\left [\left \{ \frac{1}{2}, \frac{1 -
\beta}{2} \right\}, \{1\}, k^2 \right ] k^{1 - \beta} \label{eq:
phiexpcorr}
\end{eqnarray}
with $\eta = R/R_d$ and $\eta_c = r_c/R_d$ and $k$ is still given
by Eq.(\ref{eq: defk}) replacing $(R, R', z)$ with $(\eta, \eta',
z/R_d)$. The rotation curve for the disk may be easily computed
starting from the usual relation \cite{BT87}\,:

\begin{equation}
v_c^2(R) = \left . R \frac{\partial \Phi(R, z)}{\partial R} \right
|_{z = 0} = \left . \eta \frac{\partial \Phi(R, z)}{\partial \eta}
\right |_{z/R_d = 0} \ . \label{eq: vcgen}
\end{equation}
Inserting the total gravitational potential into Eq.(\ref{eq:
vcgen}), we may still split the rotation curve in two terms as\,:

\begin{equation}
v_c^2(R) = \frac{v_{c, N}^2(R) + v_{c, corr}^2(R)}{2}
\end{equation}
where the first term is the Newtonian one, which for an
exponential disk reads \cite{Freeman70}\,:

\begin{eqnarray}
v_{c, N}^2(R) & = & 2 \pi G \Sigma_0 R_d (\eta/2)^2 \nonumber
\\
~ & {\times} & \left [ I_0(\eta/2) K_0(\eta/2) - I_1(\eta/2)
K_1(\eta/2) \right ] \label{eq: vcdisknewt}
\end{eqnarray}
with $I_l, K_l$ Bessel functions of order $l$ of the first and
second type respectively. The correction term $v_{c, corr}^2$ may
be evaluated inserting Eq.(\ref{eq: phiexpcorr}) into Eq.(\ref{eq:
vcgen}). Using\,:

\begin{displaymath}
\frac{\partial k}{\partial \eta} = \frac{k}{2 \eta} \left [ 1 -
\frac{k^2 ( \eta + \eta' )}{2 \eta'} \right ] \ ,
\end{displaymath}
we finally get\,:

\begin{equation}
v_{c, corr}^2(\eta) = - 2^{\beta - 5} \eta_c^{-\beta} \pi \alpha
(\beta - 1) G \Sigma_0 R_d \eta^{\frac{\beta - 1}{2}}
{\cal{I}}_{disk}(\eta, \beta) \label{eq: vcorr}
\end{equation}
where we have defined\,:

\begin{equation}
{\cal{I}}_{disk}(\eta, \beta) = \int_{0}^{\infty}{{\cal{F}}(\eta,
\eta', \beta) k^{3 - \beta} \eta'^{\frac{\beta - 1}{2}} {\rm e}^{-
\eta'} d\eta'}  \label{eq: deficorr}
\end{equation}
with\,:

\begin{eqnarray}
{\cal{F}} & = & 2 (\eta + \eta') {_2F_1}\left [\left \{
\frac{1}{2}, \frac{1 - \beta}{2} \right\}, \{1\}, k^2
\right ] \nonumber \\
~ & + & \left [ (k^2 - 2) \eta' + k^2 \eta \right ] {_2F_1}\left
[\left \{ \frac{3}{2}, \frac{3 - \beta}{2} \right\}, \{2\}, k^2
\right ] \ . \label{eq: deff}
\end{eqnarray}
The function ${\cal{I}}_{disk}(\eta, \beta)$ may not be evaluated
analytically, but it is straightforward to estimate it
numerically. Note that Eqs.(\ref{eq: vcorr}) and (\ref{eq:
deficorr}) can be easily generalized to a different surface
density by replacing the term ${\rm e}^{-\eta'}$ with
$\tilde{\Sigma}(\eta')$ and $R_d$ with $R_s$, being this latter a
typical scale radius of the system, while the function ${\cal{F}}$
remains unaltered.

\section{LSB rotation curves}

Historically, the flatness of  rotation curves of spiral galaxies
was the first and (for a long time) more convincing evidence for
the existence of dark matter \cite{SR01}. Despite much effort,
however, it is still unclear to what extent bright spiral galaxies
may give clues about the properties of the putative dark haloes.
On the one hand, being poor in gas content, their rotation curves
is hardly measured out to very large radii beyond the optical edge
of the disk where dark matter is supposed to dominate the rotation
curve. On the other hand, the presence of extended spiral arms and
barred structures may lead to significative non-circular motions
thus complicating the interpretation of the data. On the contrary,
LSB and dwarf galaxies are supposed to be dark matter dominated at
all radii so that the details of the visible matter distribution
are less important. In particular, LSB galaxies have an unusually
high gas content, representing up to $90\%$ of their baryonic
content \cite{vdH00,sme01}, which makes it possible to measure the
rotation curve well beyond the optical radius $R_{opt} \simeq 3.2
R_d$. Moreover, combining 21\,-\,cm HI lines and optical emission
lines such as H$\alpha$ and $[$NII$]$ makes it possible to correct
for possible systematic errors due to beam smearing in the radio.
As a result, LSB rotation curves are nowadays considered a useful
tool to put severe constraints on the properties of the dark
matter haloes (see, e.g., de Blok 2005 and references therein).

\begin{table}
\caption{Properties of sample galaxies. Explanation of the
columns\,: name of the galaxy, distance in Mpc; disk central
surface brightness in the $R$ band (corrected for galactic
extinction); disk scalelength in kpc; radius at which the gas
surface density equals $1 \ {\rm M_{\odot}/pc^2}$ in arcsec; total
HI gas mass in $10^8 \ {\rm M_{\odot}}$; Hubble type as reported
in the NED database.}
\begin{center}
\begin{tabular}{|c|c|c|c|c|c|c|}
\hline
Id & $D$ & $\mu_0$ & $R_d$ & $R_{HI}$ & $M_{HI}$ & Type \\
\hline UGC 1230 & 51 & 22.6 & 4.5 & 101 & 58.0 & Sm\\
UGC 1281 & 5.5 & 22.7 & 1.7 & 206 & 3.2 & Sdm\\ UGC 3137 & 18.4 &
23.2 & 2.0 & 297 & 43.6 & Sbc \\ UGC 3371 & 12.8 & 23.3 & 3.1 & 188 & 12.2 & Im \\
UGC 4173 & 16.8 & 24.3 & 4.5 & 178 & 21.2 & Im \\ UGC 4325 & 10.1
& 21.6 & 1.6 & 142 & 7.5 & SAm \\ NGC 2366 & 3.4 & 22.6 & 1.5 &
439 & 7.3 & IB(s)m \\ IC 2233 & 10.5 & 22.5 & 2.3 & 193 & 13.6 &
SBd \\ NGC 3274 & 6.7 & 20.2 & 0.5 & 225 & 6.6 & SABd \\ NGC 4395
& 3.5 & 22.2 & 2.3 & 527 & 9.7 & SAm \\ NGC 4455 & 6.8 & 20.8 &
0.7 & 192 & 5.4 & SBd \\ NGC 5023 & 4.8 & 20.9 & 0.8 & 256 & 3.5 &
Scd \\ DDO 185 & 5.1 & 23.2 & 1.2 & 136 & 1.6 & IBm \\ DDO 189 &
12.6 & 22.6 & 1.2 & 167 & 10.5 & Im \\ UGC 10310 & 15.6 & 22.0 &
1.9 & 130 & 12.6 & SBm \\ \hline
\end{tabular}
\end{center}
\end{table}

\subsection{The data}

It is easy to understand why LSB rotation curves are ideal tools
to test also modified gravity theories. Indeed, successfully
fitting the rotation curves of a whatever dark matter dominated
system, without resorting to dark matter, should represent a
serious evidence arguing in favour of modifications of the
standard Newtonian potential. In order to test our model, we have
therefore considered a sample of 15 LSB galaxies with well
measured HI and H$\alpha$ rotation curves extracted from a larger
sample in de Blok \& Bosma (2002). The initial sample contains 26
galaxies, but we have only considered those galaxies for which
data on the rotation curves, the surface photometry in the $R$
band and the gas mass surface density were available\footnote{This
initial selection reduced indeed the sample to 19 galaxies, but
four of them were rejected because of numerical problems when
computing the gas rotation curve due to the strong irregularities
in the interpolated surface density.}. In Table 1, we report the
quantities we need for evaluating the theoretical rotation curve
referring the reader to de Blok \& Bosma (2002) for further
details and references to retrieve the data\footnote{The data on
the rotation curves may be also found in the SIMBAD database ({\tt
http://cdsweb.u-strasbg.fr}).}.

An important remark in in order here. For each LSB galaxy, both HI
and H$\alpha$ data on the rotation curve are available. As yet
discussed in de Blok \& Bosma (2002), the raw data show some
scatter mainly due to residuals non circular motions that may lead
to ambiguous rotational velocities. However, when deriving mass
models from rotation curves, each galaxy is described as an
axisymmetric system so that non circular motions do not arise. In
order to remove this scatter from the data, it is recommended to
use the smoothed rotation curve data derived by a local regression
method extensively discussed in de Blok \& Bosma (2002 and refs.
therein). Following these authors, we will adopt the smooth data
as input in the fitting procedure. The smoothing procedure may in
principle introduce correlations among the data so that it is
worth investigating whether this may bias somewhat the results on
the model parameters. Moreover, the number of data points on each
single rotation curve is reduced and the errors on each point is
estimated in a different way than for raw data. As such, it is
important to investigate also how this affects the uncertainties
on the final estimate of the model parameters.

\subsection{Modelling LSB galaxies}

Since we are interested in fitting rotation curves without any
dark matter halo, our model for a generic LSB galaxy is made out
of the stellar and gaseous components only.

We assume the stars are distributed in an infinitely thin and
circularly symmetric disk. The surface density $\Sigma(R)$ may be
derived from the surface brightness distribution\,:

\begin{displaymath}
\mu(R) = -2.5 \log{I(R)}
\end{displaymath}
with $I(R) = \Sigma(R)/\Upsilon_{\star}$ the light distribution
and $\Upsilon_{\star}$ the stellar mass\,-\,to\,-\,light
(hereafter $M/L$) ratio. The photometric data (in the $R$ band)
are fitted with an exponential model thus allowing to determine
the scalelength $R_d$ and the central surface brightness $\mu_0$
and hence $I_0 = I(R = 0)$. The only unknown parameter is
therefore $\Upsilon_{\star}$ that makes it possible to convert the
central luminosity density $I_0$ into the central surface mass
density $\Sigma_0$ entering Eqs.(\ref{eq: vcdisknewt}) and
(\ref{eq: vcorr}).

Modelling the gas distribution is quite complicated. Following the
standard practice, we assume the gas is distributed in a
infinitely thin and circularly symmetric disk assuming for the
surface density $\Sigma(R)$ the profile that has been measured by
the HI 21\,-\,cm lines. Since the measurements only cover the
range $R_{min} \le R \le R_{max}$, we use a third order
interpolation for $R$ in this range, a linear extrapolation
between $R_{max}$ and $R_{HI}$, being this latter a scaling radius
defined by $\Sigma(R_{HI}) = 1 \ {\rm M_{\odot}/pc^2}$, while we
assume $\Sigma(R) = \Sigma(R_{min})$ for $R \le R_{min}$.  To
check if the model works correctly, we compute the total mass
$M_{HI}$ and normalize the model in such a way that this value is
the same as that is measured by the total HI 21\,-\,cm emission.
Finally, we increase the surface mass density by 1.4 to take into
account the helium contribution. It is worth noting that our model
is only a crude approximation for $R$ outside the range $(R_{min},
R_{max})$, while, even in the range $(R_{min}, R_{max})$,
$\Sigma(R)$ gives only an approximated description of the gas
distribution since this latter may be quite clumpy and therefore
cannot be properly fitted by any analytical expression. We stress,
however, that the details of the gas distribution are rather
unimportant since the rotation curve is dominated everywhere by
the stellar disk. The clumpiness of the gas distribution manifests
itself in irregularities in the rotation curve that may be easily
masked in the fitting procedure, even if this is not strictly
needed for our aims.

\begin{figure*}
\centering
\includegraphics[width=16cm]{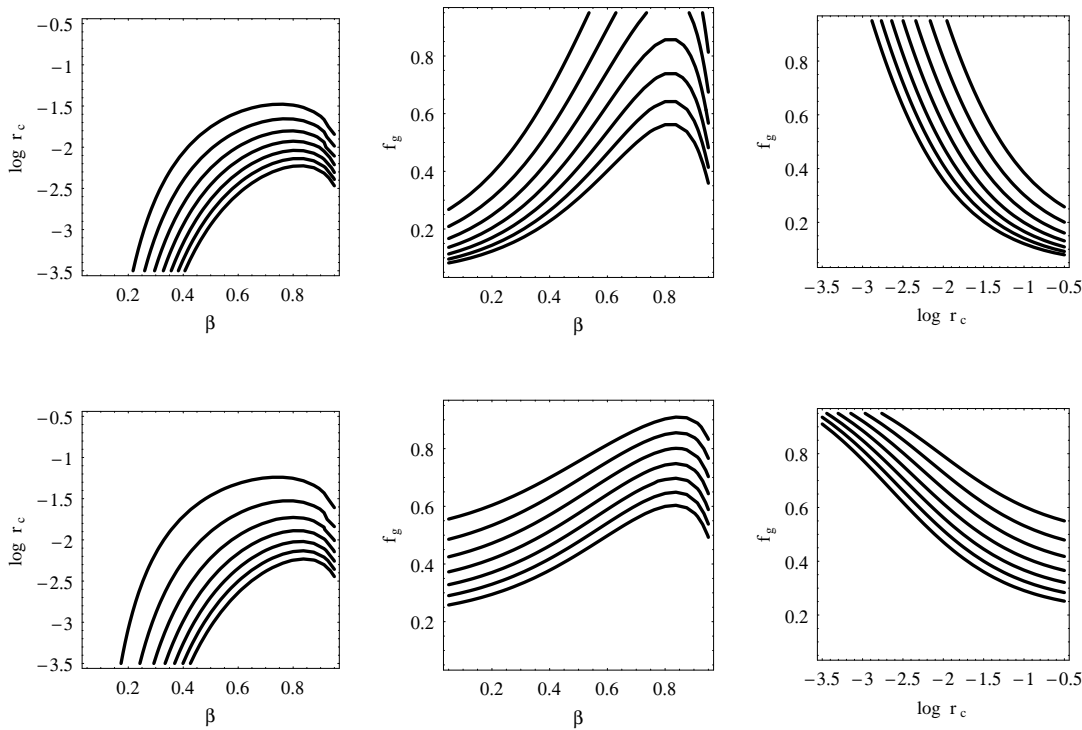}
\caption{Contour plots for $v_c(R_d)$ in the planes $(\beta,
\log{r_c})$, (left), $(\beta, f_g)$ (middle), $(\log{r_c}, f_g)$
(right) with $r_c$ in kpc. The contours are plotted for $v_c(R) =
k \times v_{fid}$ with $k$ from 0.7 to 1.3 in steps of 0.1 and
$v_{fid} = v_c(R_d)$ for the model with $(\beta, \log{r_c}, f_g) =
(0.61, -2.13, 0.65)$. Upper panels refer to a pointlike system
with total mass $m = \Upsilon_{\star} L_d + M_{HI}$, with $L_d$
the total disk luminosity, $M_{HI}$ the gas mass and
$\Upsilon_{\star}$ given by Eq.(\ref{eq: fgdef}). Lower panels
refer to the extended case using as default parameters those of
UGC 10310. In each panel, the remaining parameter is set to its
fiducial value. Note that similar plots are obtained for values of
$R$ other than $R_d$.} \label{fig: vcont}
\end{figure*}

\subsection{Fitting the rotation curve}

Having modelled a LSB galaxy, Eqs.(\ref{eq:
vcdisknewt})\,--\,(\ref{eq: deff}) may be straightforwardly used
to estimate the theoretical rotation curve as function of three
unknown quantities, namely the stellar $M/L$ ratio
$\Upsilon_{\star}$ and the two theory parameters $(\beta, r_c)$.
Actually, we will consider as fitting parameters $\log{r_c}$
rather than $r_c$ (in kpc) since this is a more manageable
quantity that makes it possible to explore a larger range for this
theoretically unconstrained parameter. Moreover, we use the gas
mass fraction $f_g$ rather than $\Upsilon_{\star}$ as fitting
quantity since the range for $f_g$ is clearly defined, while this
is not for $\Upsilon_{\star}$. The two quantities are easily
related as follows\,:

\begin{equation}
f_g = \frac{M_{g}}{M_{g} + M_d} \iff \Upsilon_{\star} = \frac{(1 -
f_g) M_{g}}{f_g L_d} \label{eq: fgdef}
\end{equation}
with $M_g = 1.4 M_{HI}$ the gas (HI + He) mass, $M_d =
\Upsilon_{\star} L_d$ and $L_d = 2 \pi I_0 R_d^2$ the disk total
mass and luminosity.

We use Eq.(\ref{eq: vcdisknewt}) to compute the disk Newtonian
rotation curve, while the $v_{c,corr}$ is obtained by integrating
numerically Eq.(\ref{eq: deficorr}). For the gas, instead, we
resort to numerical integrations for both the Newtonian rotation
curve and the corrective term. The total rotation curve is finally
obtained by adding in quadrature these contributions.

To constrain the parameters $(\beta, \log{r_c}, f_g)$, we minimize
the following merit function\,:

\begin{equation}
\chi^2({\bf p}) = \sum_{i = 1}^{N}{\left [ \frac{v_{c,th}(r_i) -
v_{c,obs}(r_i)}{\sigma_i} \right ]^2} \label{eq: defchi}
\end{equation}
where the sum is over the $N$ observed points. While using the
smoothed data helps in better adjusting the theoretical and
observed rotation curves, the smoothing procedure implies that the
errors $\sigma_i$ on each point are not Gaussian distributed since
they also take into account systematic misalignments between HI
and H$\alpha$ measurements and other effects leading to a
conservative overestimate of the true uncertainties (see the
discussion in \cite{dbb02} for further details). As a consequence,
we do not expect that $\chi^2/dof \simeq 1$ for the best fit model
(with $dof = N - 3$ the number of degrees of freedom), but we can
still compare different models on the basis of the $\chi^2$
values. In particular, the uncertainties on the model parameters
will be estimated exploring the contours of equal $\Delta \chi^2 =
\chi^2 - \chi^2_{min}$ in the parameter space.

\section{Testing the method}

The method we have outlined above and the data on LSB galaxies are
in principle all what we need to test the viability of $R^n$
gravity. However, there are some subtle issues that can affect in
an unpredictable way the outcome of the analysis.

\begin{figure*}
\centering \subfigure{\includegraphics[width=8cm]{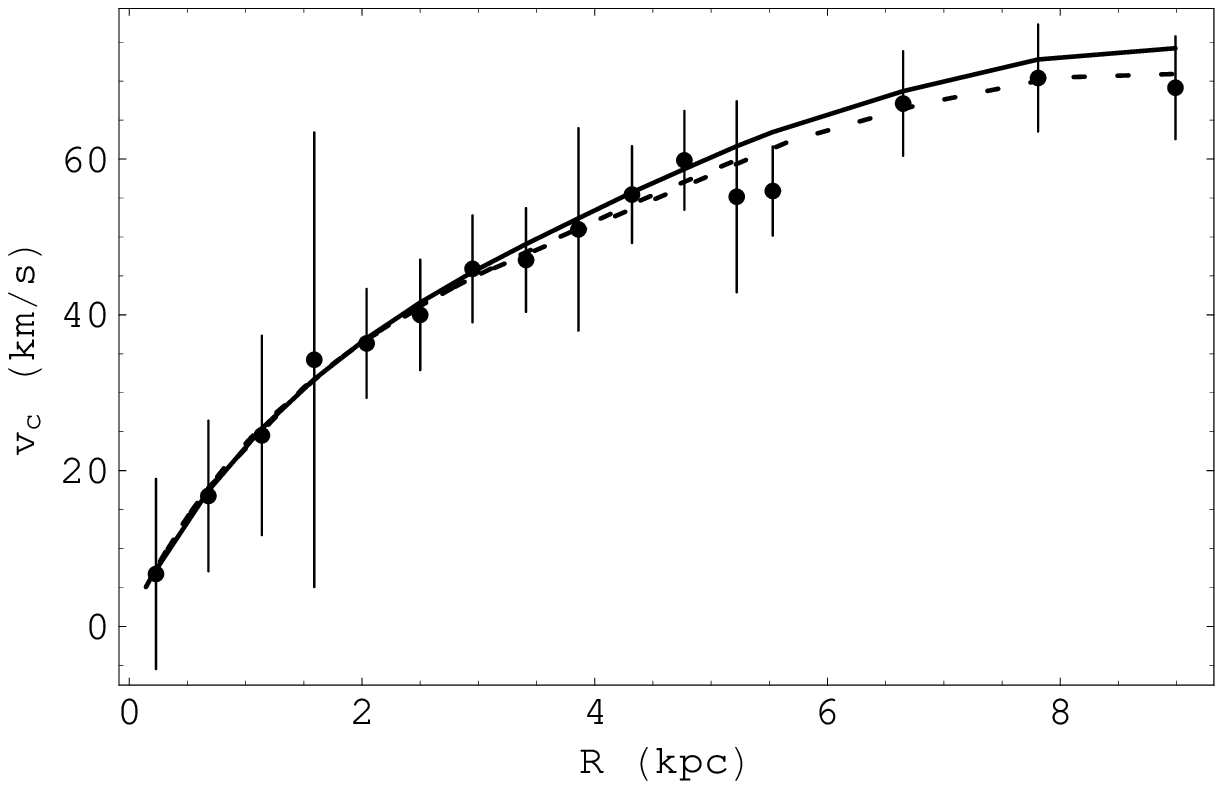}}
\goodgap \subfigure{\includegraphics[width=8cm]{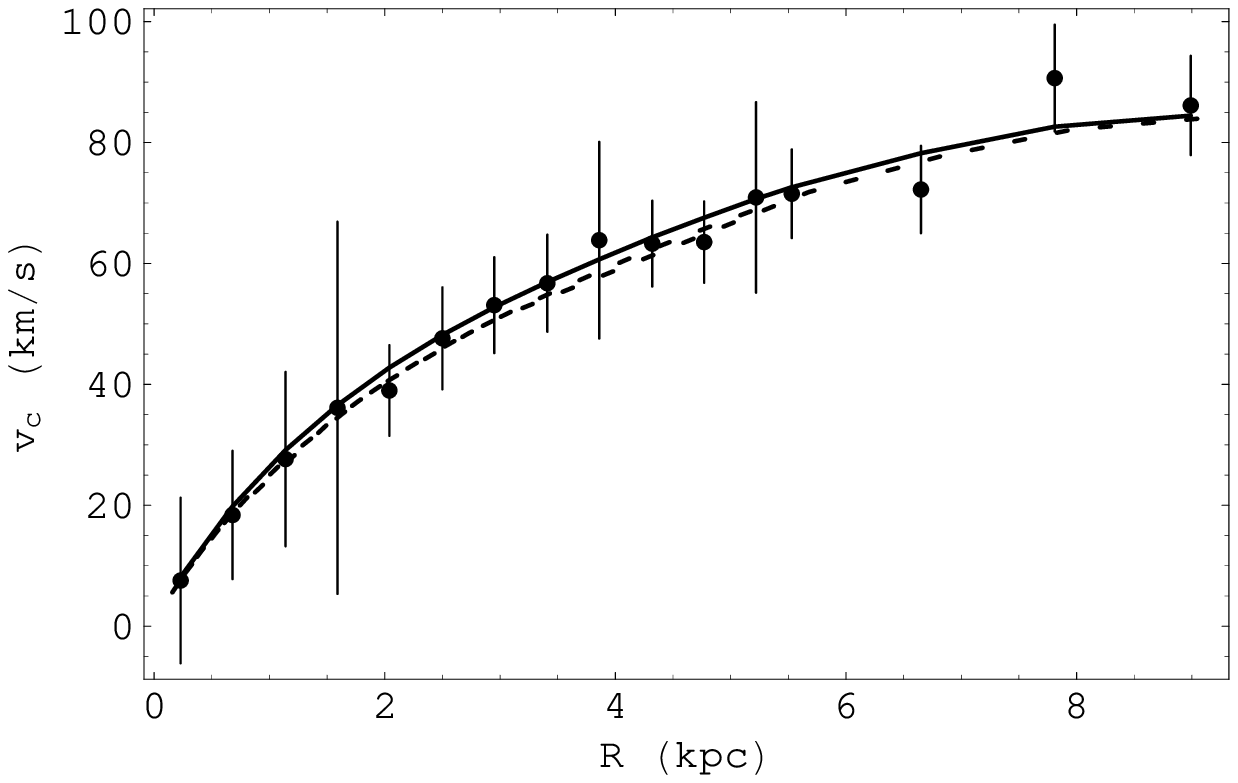}} \\
\subfigure{\includegraphics[width=8cm]{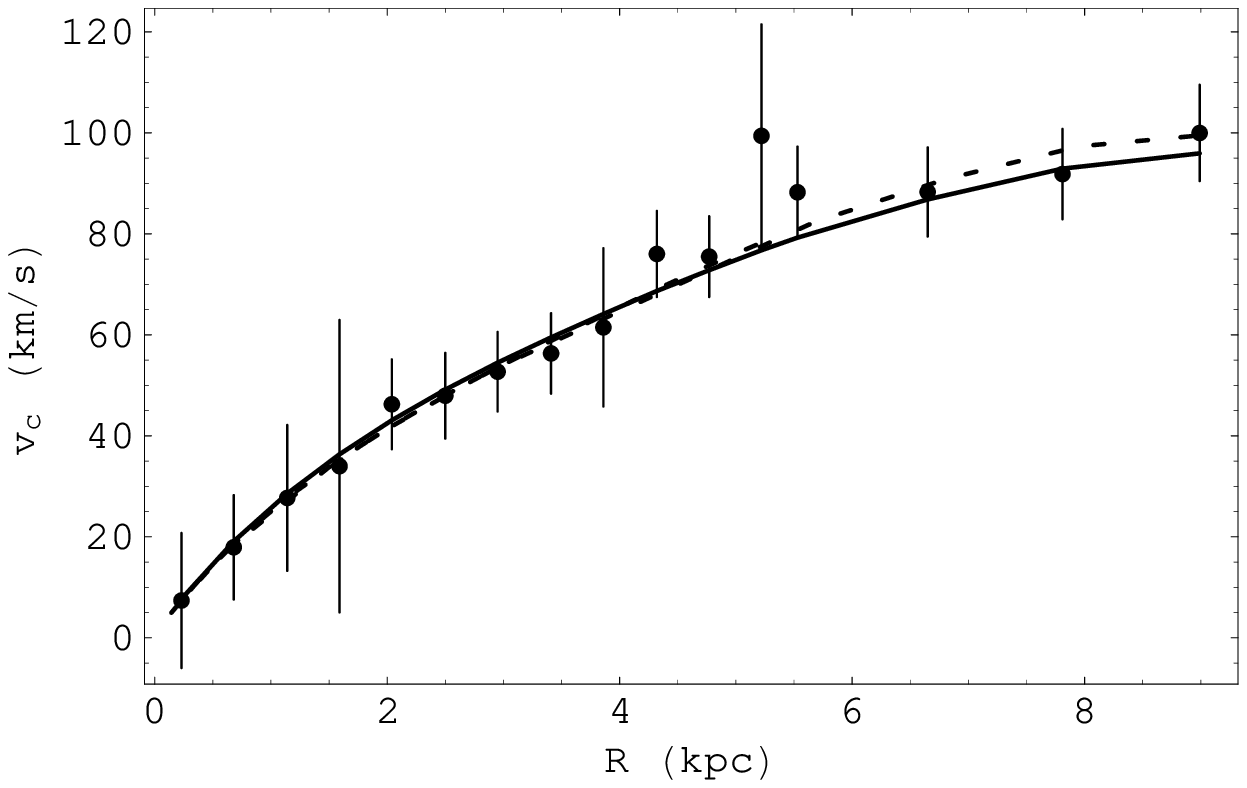}} \goodgap
\subfigure{\includegraphics[width=8cm]{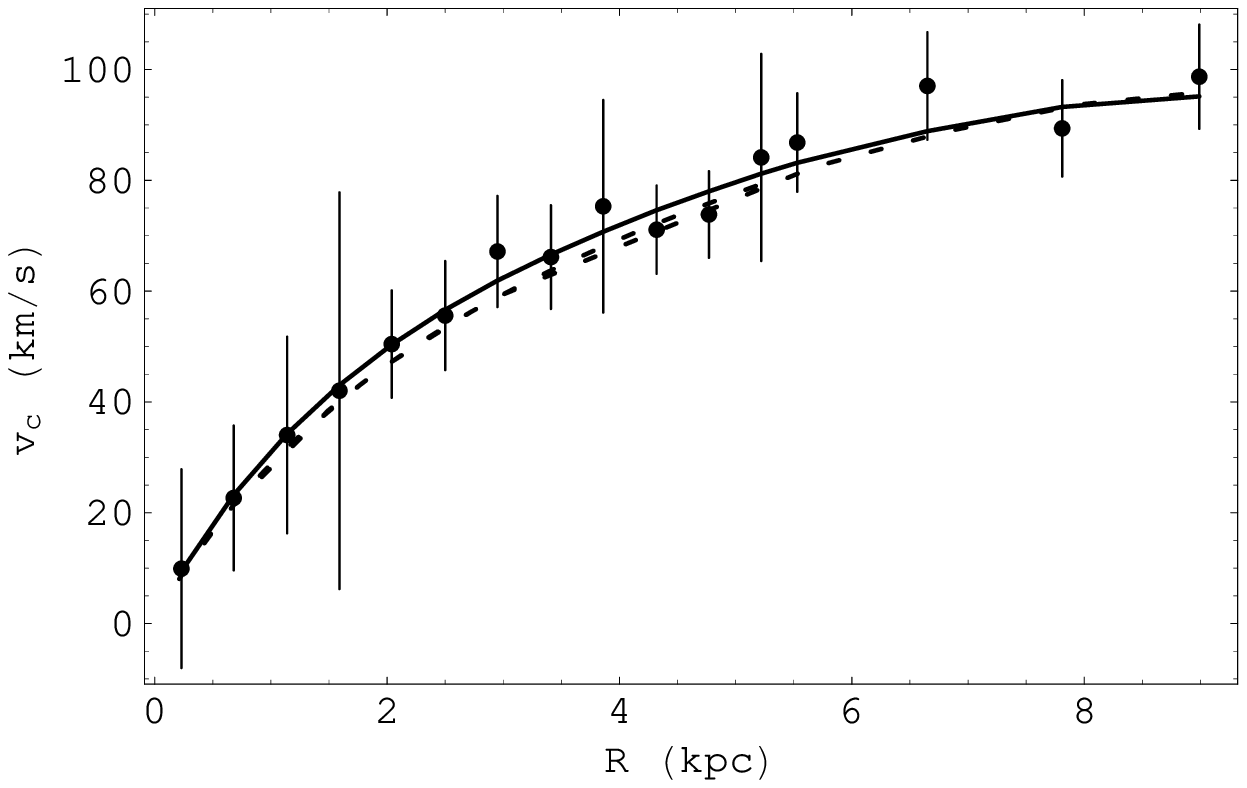}} \goodgap \\
\caption{Some illustrative examples of simulated rotation curves
(smoothing the data for convenience) with overplotted the input
theoretical rotation curve (solid line) and the best fit one
(short dashed line).}
\label{fig: simthree}
\end{figure*}

Two main problems worth to be addressed. First, there are three
parameters to be constrained, namely the gas mass fraction $f_g$
(related to the stellar $M/L$ ratio $\Upsilon_{\star}$) and the
$R^n$ gravity quantities $(\beta, \log{r_c})$. However, although
they do not affect the theoretical rotation curve in the same way,
there are still some remaining degeneracies hard to be broken.
This problem is well illustrated by Fig.,\ref{fig: vcont} where we
show the contours of equal $v_c(R_d)$ in the planes $(\beta,
\log{r_c})$, $(\beta, f_g)$ and $(\log{r_c}, f_g)$ for the
pointlike and extended case. Looking, for instance, at right
panels, one sees that, for a given $\beta$, $\log{r_c}$ and $f_g$
(and hence $\Upsilon_\star$) have the same net effect on the
rotation curve so that the same value for $v_c(R_d)$ may be
obtained for a lower $f_g$ provided one increases $\log{r_c}$. On
the other hand, $\beta$ and $\log{r_c}$ have opposite effects on
$v_c(R)$\,: the lower is $\beta$, the smaller is $v_c(R)$ for a
given $R$. Since the opposite holds for $\log{r_c}$, as a result,
the same value of $v_c(R_d)$ may be obtained increasing
$\log{r_c}$ and decreasing $\beta$. Moreover, while $\beta$ drives
the shape of the rotation curve in the outer region, its effect
may be better appreciated if $r_c$ is low so that a further
degeneracy arises.

A second issue is related to our decision to use the smooth rather
than the raw data. Although de Blok \& Bosma (2002) claim that
this does not affect the results, their analysis is nevertheless
performed in the framework of standard theory of gravity with dark
matter haloes. It is therefore worth investigating whether this
holds also in the case of the $R^n$ gravity we are considering
here.

Both these issues may be better investigated through the analysis
of simulated rotation curves. To this aim, we take UGC 10310 as
input model for the gas surface density and the disk luminosity
because its properties are typical of our sample. For given values
of the model parameters $(\beta, \log{r_c}, f_g)$, we generate
observed rotation curves using the same radial sampling of the
actual observations. For each $r_i$, we randomly extract
$v_{c,obs}(r_i)$ from a Gaussian distribution centred on the
$v_{c,th}(r_i)$. To this point, we attach an error extracted from
a second Gaussian distribution centred on the $\varepsilon_i
{\times} v_{c,obs}(r_i)$ with $\varepsilon_i$ the percentage error
on $v_{c,obs}(r_i)$ in the real sample. The simulated rotation
curves thus obtained are quite similar to the raw data so that we
use the same smoothing procedure (as in Appendix B) to get
simulated smooth data. Both the simulated raw and smooth data are
quite similar to the corresponding observed ones so that they
represent ideal tools to explore the issues quoted above.

\subsection{The impact of the parameters degeneracy}

As well known, the determination of ${\cal{N}}$ model parameters
from the fit to a given dataset may be seriously compromised if
strong degeneracies exist. Considering for simplicity the case of
a pointlike source, the rotation curve may be roughly approximated
as\,:

\begin{equation}
v_c^2(r) \simeq \frac{G m}{2 r} {\times} \left \{
\begin{array}{ll}
1 & {\rm for \ } (r/r_c)^{\beta} << 1 \\
 1 + (1 - \beta) & {\rm for \ } (r/r_c)^{\beta}
\simeq 1 \\ (1 - \beta) (r/r_c)^{\beta} & {\rm for \ }
(r/r_c)^{\beta} >> 1
\end{array}
\right . \ \nonumber . \label{eq: vcpointapp}
\end{equation}
For the typical values of $\beta$ ($\sim 0.8$) and $r_c$ ($\simeq
0.01 \ {\rm kpc}$) we qualitatively estimate from an eyeball fit
to the data, it is easy to check that most of the data in the
rotation curves mainly probe the region with $(r/r_c)^{\beta} >>
1$ so that $v_c^2 \simeq Gm/2 r_c {\times} (1 - \beta)
(r/r_c)^{\beta - 1}$. As a result, the theoretical rotation curve
mainly depends on the two effective parameters $m_{eff} = m (1 -
\beta)/2$ and $r_{c,eff} = r_c^{\beta}$. Moreover, as a further
complication, the lower is $\beta$, the less $v_c^2$ depends on
$r_c$ since the correction term is more and more negligible. A
similar discussion (although less intuitively) also holds in the
extended case with the stellar $M/L$ ratio $\Upsilon_{\star}$
playing the role of the pointlike mass $m$. Both these problems
may be better appreciated looking at Fig.\,\ref{fig: vcont} as yet
discussed above. It is therefore mandatory to explore whether the
data are able to break these degeneracies or how they affect the
recovering of the model parameters $(\beta, \log{r_c}, f_g)$.

\begin{figure*}
\centering
\includegraphics[width=16cm]{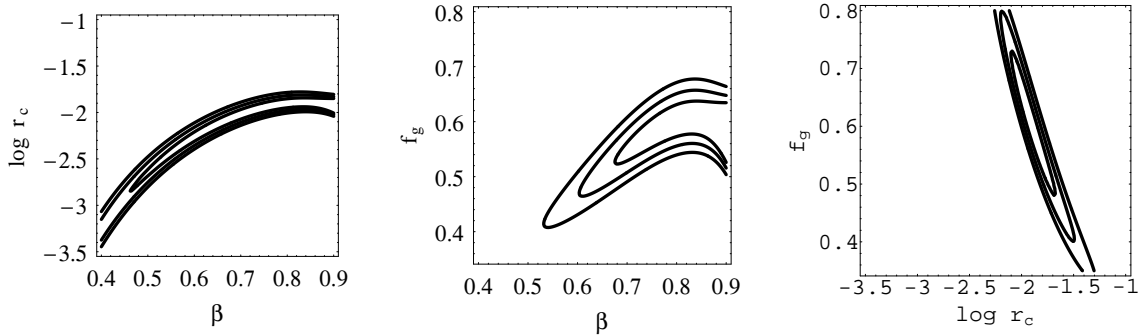}
\caption{Contours of equal $\Delta \chi^2 = \chi^2 - \chi^2_{min}$
projected on the three planes $(\beta, \log{r_c})$, $(\beta,
f_g)$, $(\log{r_c}, f_g)$ for the case of the simulation in the
top right panel of Fig. \ref{fig: simthree} with $r_c$ in kpc. In
each panel, the remaining parameter is set to its best fit value.
The three contours individuate the $1$, $2$ and $3 \sigma$
confidence ranges. Open contours mean that no constraints may be
obtained.} \label{fig: s3e5cl}
\end{figure*}

This task may be ideally tackled fitting the simulated rotation
curves generated as described above and comparing the best fit
values with the input ones. Indeed, we find that the degeneracy
works in a quite dramatic way possibly leading to large
discrepancies among the input and best fit values. To better
illustrate this point, some typical examples have been reported in
Fig. \ref{fig: simthree} where the input theoretical rotation
curve (solid line) and the best fit one (short dashed line) are
superimposed to the simulated data. Note that we plot smoothed
rather than raw data in order to not clutter the graphic with too
many points, but the raw data have been used in the fit. Although
the two lines in each panel are always remarkably close so that
they can be hardly discriminated by the data, the offset $\Delta
p/p = |1 - p_{fit}/p_{sim}|$ may be quite large. Considering, for
instance, the parameter $\beta$, we get $\Delta \beta /\beta =
24\%$, $11\%$, $17\%$ and $16\%$ from top left to bottom right
clockwise respectively. Similar results are obtained for the full
set of simulations, while smaller values of $\Delta p/p$ come out
for $p = \log{r_c}$ and $p = f_g$ (and hence for
$\Upsilon_{\star})$. It is also worth stressing that no
significant correlation has been observed between $\Delta p/p$ and
$p$ whatever is the parameter $p$ considered.

This exercise also teaches us an important lesson. As it is
apparent, better quality data could not be sufficient to break the
degeneracy. An instructive example is represented by the top right
panel where the two curves almost perfectly overlap. It is clear
that reducing the error bars does not help at all so the input and
the best fit models are impossible to discriminate notwithstanding
a remarkable $\Delta \beta/\beta = 11\%$. In some cases, the two
curves start departing from each other for large $R$ so that one
could expect that adding more points in this region or extending
the data to still higher $R$ efficiently breaks the degeneracy.
Unfortunately, the simulated data extend up to $\sim 5 R_d$ so
that further increasing this coverage with real data is somewhat
unrealistic (especially using typical spiral galaxies rather than
the gas rich LSBs).

The degeneracy hinted above among the model parameters also
dramatically affects the estimated errors on $(\beta, \log{r_c},
f_g)$. This can be seen in Fig. \ref{fig: s3e5cl} where we report
the contours of equal $\Delta \chi^2$ projected on the three
planes $(\beta, \log{r_c})$, $(\beta, f_g)$, $(\log{r_c}, f_g)$
for the case of the simulation in the top right panel of Fig.
\ref{fig: simthree}. As it is clearly seen, $\beta$ remains
essentially unconstrained, while $\log{r_c}$ and $f_g$ are only
weakly constrained with the $3 \sigma$ contours still spanning
almost the full physical range for $f_g$. As suggested by the
analysis of the pointlike case, these discouraging results are an
expected consequence of the dependence of the theoretical rotation
curve on the two degenerate quantities $(m_{eff}, r_{c,eff})$
preventing us to efficiently constraint separately the three model
parameters.

\begin{figure*}
\centering
\subfigure{\includegraphics[width=8cm]{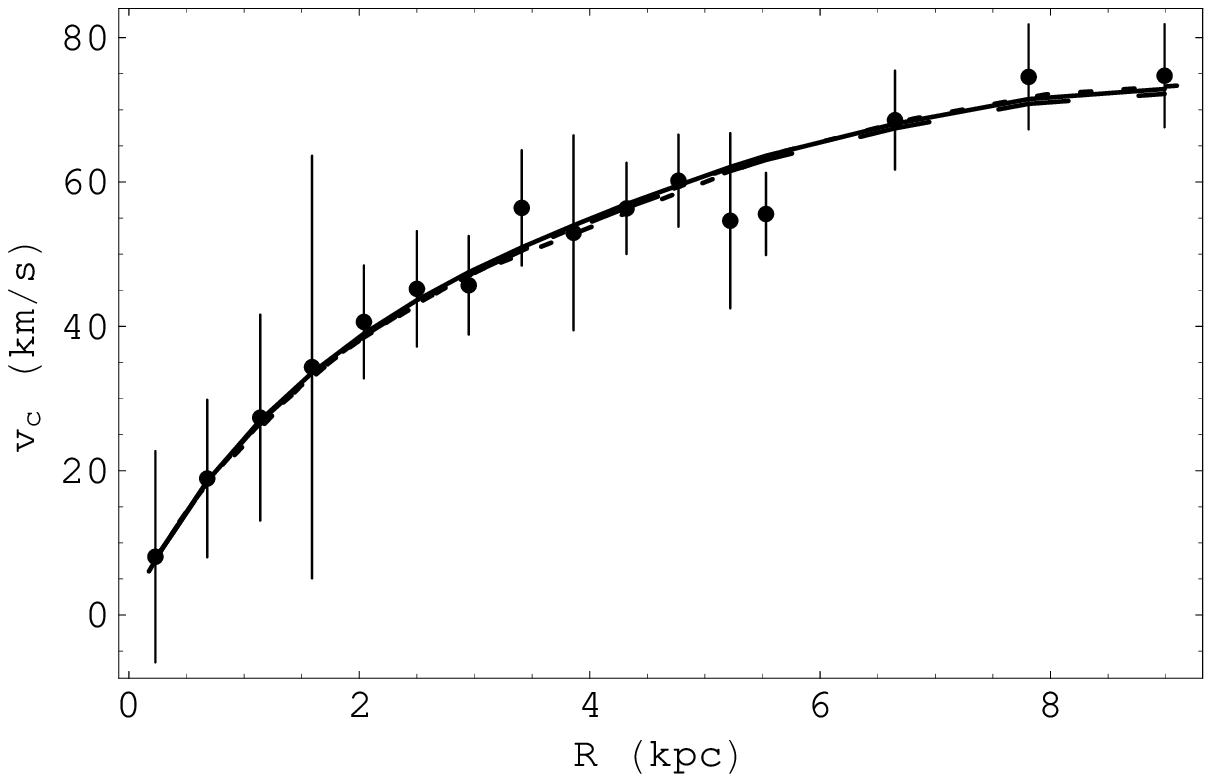}}
\goodgap \subfigure{\includegraphics[width=8cm]{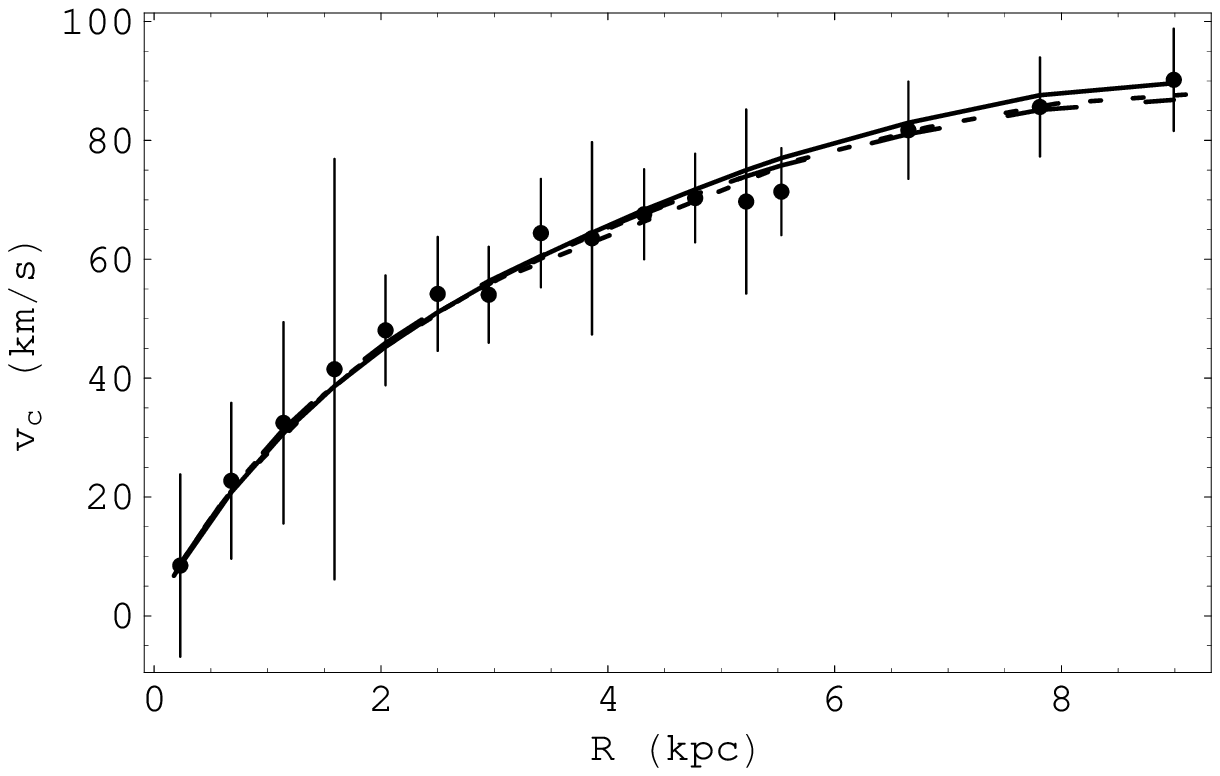}} \\
\subfigure{\includegraphics[width=8cm]{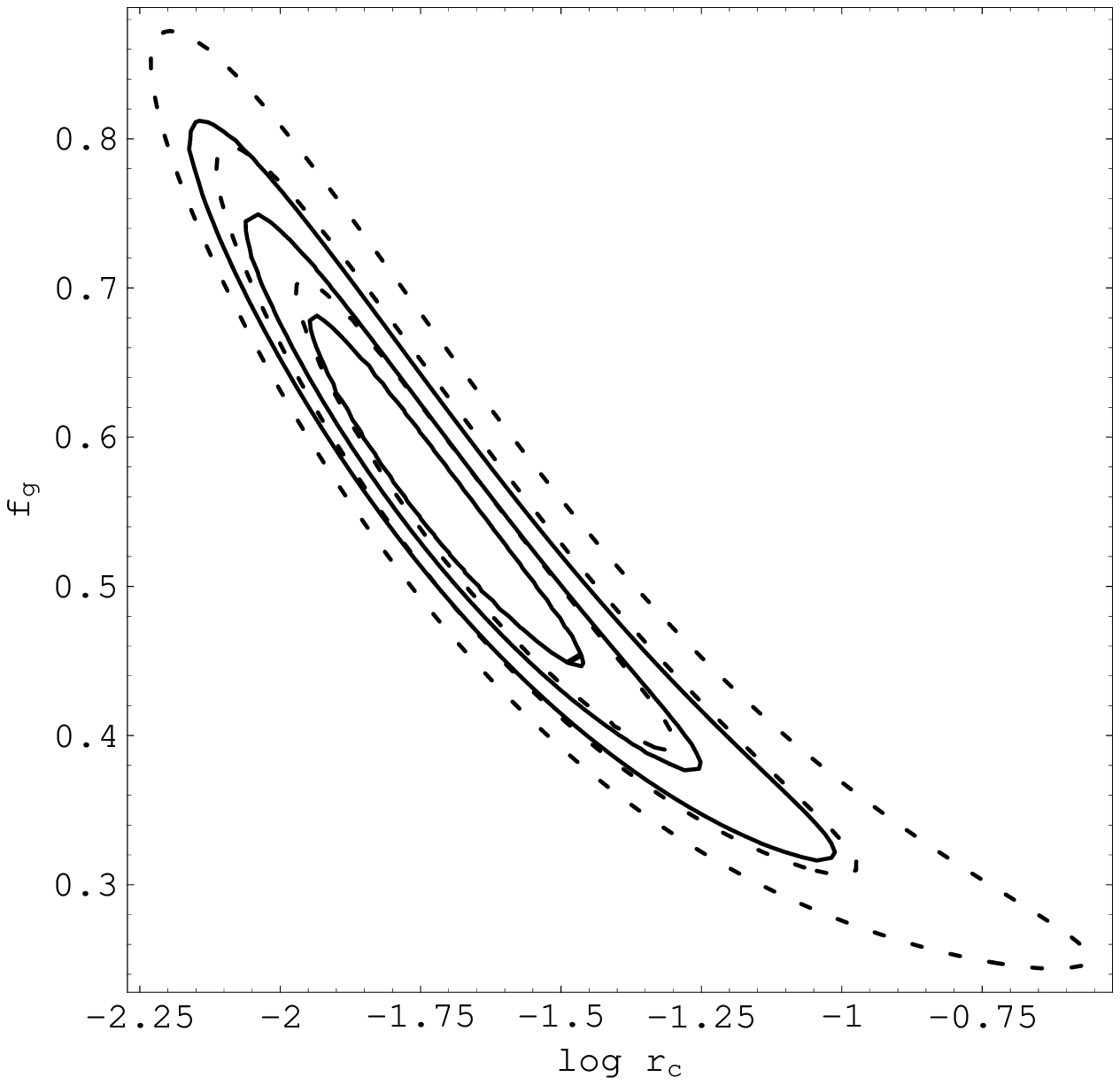}} \goodgap
\subfigure{\includegraphics[width=8cm]{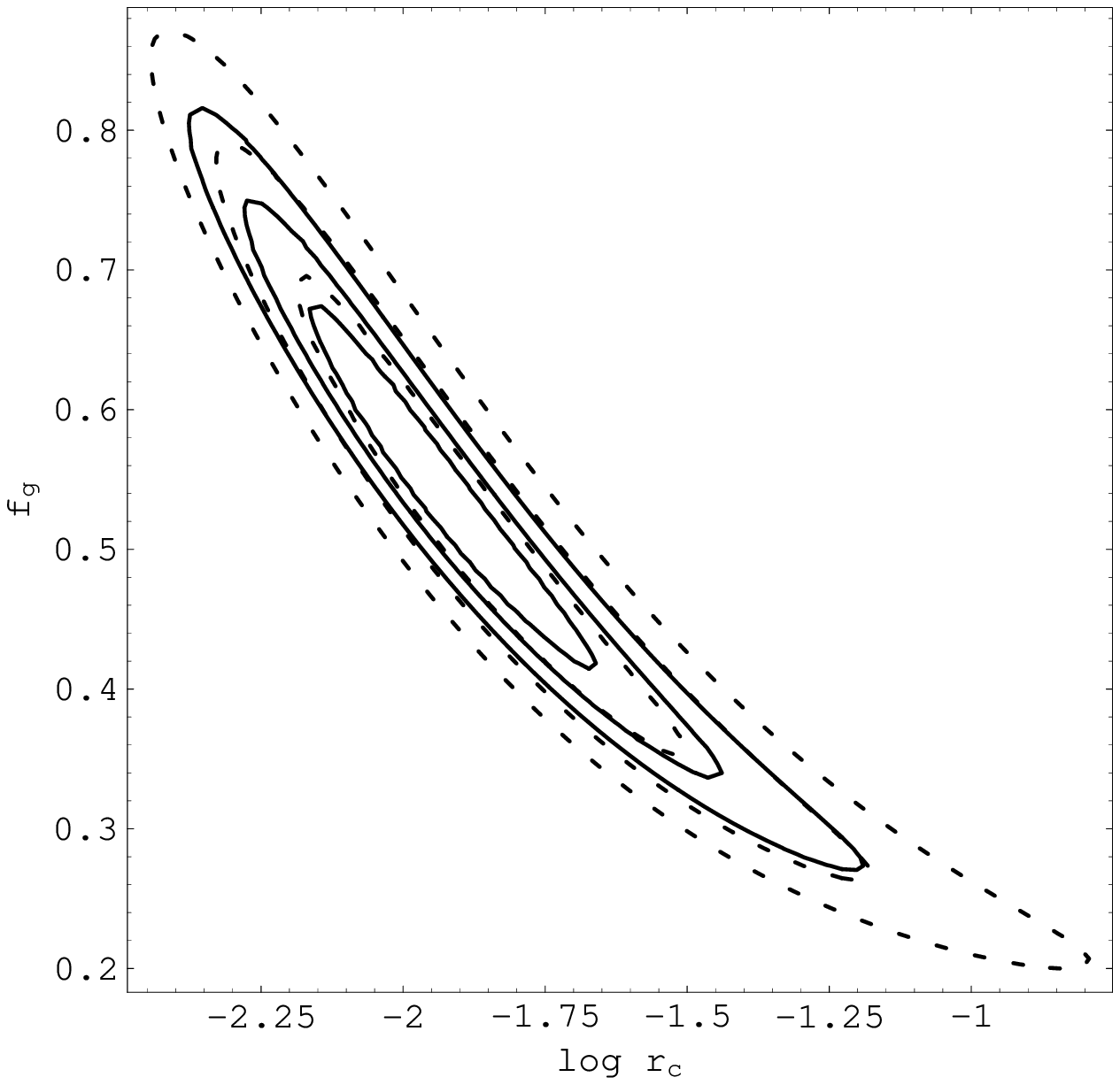}} \goodgap \\
\caption{{\it Top panels.} Some illustrative examples of simulated
rotation curves (smoothing the data for convenience) with
overplotted the input theoretical rotation curve (solid line) and
the best fit one from raw (short dashed line) and smooth (long
dashed line) data. {\it Bottom panels.} $1$, $2$ and $3 \sigma$
confidence ranges in the plane $(\log{r_c}, f_g)$ from the fit to
raw (solid line) and smooth (short dashed line) data shown in the
respective top panels (with $r_c$ in kpc). Note that the two cases
reported are representatives of the best (left) and worst (right)
situations we find in our sample of simulated rotation curves.}
\label{fig: simtwo}
\end{figure*}

\subsection{Breaking the degeneracy among $(\beta, \log{r_c}, f_g)$}

The analysis carried out convincingly shows that the degeneracy
among the model parameters can not be broken by the present data
on the rotation curves alone. As a result, one has to add some
more constraints coming from different sources in order to set one
of the three parameters above thus breaking the degeneracy and
correctly recovering the values of the remaining two. Again, the
analysis of the simulated rotation curves may help in choosing the
best strategy.

As a first possibility, one may resort to stellar population
synthesis models in order to set the $M/L$ ratio
$\Upsilon_{\star}$ and hence the gas mass fraction $f_g$. Although
this strategy is not free of problems (see the discussion in Sect.
6), one can ideally correlate the observed colors of the galaxy to
the predicted $M/L$ ratio and hence performing the fit to the data
with the parameter $f_g$ obtained by Eq.(\ref{eq: fgdef}) so that
$(\beta, \log{r_c})$ are the only unknown quantities. The fit
results to the full set of simulated rotation curves unequivocally
show us that this strategy does not work at all. Although we do
not report any illustrative examples, we warn the reader that
quite similar plots to Fig. \ref{fig: simthree} are obtained.
Actually, while $|\Delta \log{r_c}/\log{r_c}|$ is reduced to $\sim
10\%$, we still have $|\Delta \beta/\beta| \sim 20\%$ with values
as high as $40\%$. Nevertheless, the input theoretical and the
best fit rotation curves overlaps quite well over the range probed
by the data.

As in the case with all the parameters free to vary, reducing the
errors bars or extending the radial coverage is typically not
sufficient for lowering $\Delta \beta/\beta$. Such a result may be
anticipated by considering again Eq.(\ref{eq: vcpointapp}).
Setting $\Upsilon_{\star}$ is the same as choosing $m$ so that one
could argue that $(\beta, \log{r_c})$ may be determined by the
effective quantities $(m_{eff}, r_{c,eff})$ that are constrained
by the data. Actually, the situation is much more involved.
Indeed, for $\beta << 1$, we are in a quasi Newtonian regime so
that $r_{c, eff}$ is very weakly constrained and hence neither
$\beta$ nor $r_c$ may be recovered. On the other hand, if $\beta
\simeq 1$ the correction is small and again the constraints on
both parameters are weak.

The simple exercise discussed above shows us that also a perfect
knowledge of $\Upsilon_{\star}$ is unable to break the remaining
degeneracy between $\beta$ and $\log{r_c}$ thus preventing the fit
to recover their correct value. As a second possibility, one may
resort to the theory itself and decide to set $\beta$ from the
beginning. Actually, this is the same as setting the slope $n$ of
the $R^n$ gravity Lagrangian. Since this latter must be the same
from the galactic up to the cosmological scales, one may determine
$n$ from a different test and then set $\beta$ from Eq.(\ref{eq:
bnfinal}). The fit to the data may then be used to estimate
$(\log{r_c}, f_g)$ which, on the contrary, depend on the
particular system under examination. Indeed, we find that this
strategy works very well. Both $\log{r_c}$ and $f_g$ are recovered
with great accuracy being $|\Delta \log{r_c}/\log{r_c}| \sim
|\Delta f_g/f_g| \sim 5\%$ and never greater than $\sim 10\%$. Two
cases representative of our best and worst situations are shown in
Fig. \ref{fig: simtwo}. In both cases, the input theoretical curve
and the best fit one may be hardly distinguished and indeed we get
$(\Delta \log{r_c}/\log{r_c}, \Delta f_g/f_g) = (-2\%, -3\%)$ for
the case in the left panel and $(3\%, 5\%)$ for the one in the
right panel.

Considering the intrinsic errors induced by the displacement from
the input rotation curve induced by our procedure used to generate
the simulated data, we can safely conclude that both $\log{r_c}$
and $f_g$ are exactly recovered within the expected precision.

It is also interesting to look at the bottom panels in Fig.
\ref{fig: simtwo} showing the iso\,-\,$\Delta \chi^2$ contours in
the plane $(\log{r_c}, f_g)$. Although still covering a large
region of the parameter space, the confidence ranges are now
closed so that it is possible to extract meaningful constraints on
the parameters. Following the usual approach (see, e.g.,
\cite{dbb02}), $1$ and $2 \sigma$ errors are obtained by
projecting on the axes the contours for $\Delta \chi^2 = 1$ and
$\Delta \chi^2 = 4$ respectively\footnote{We caution the reader
that the contours in Figs. \ref{fig: s3e5cl} and \ref{fig: simtwo}
refer to 2D confidence ranges so that $\Delta \chi^2 = 2.30, 6.17,
11.8$ respectively. They do not must be confused with those
reported in the text which refer to constraints in a 1D parameter
space.}. A naive propagation of errors on $f_g$ and the use of
Eq.(\ref{eq: fgdef}) makes it possible to infer constraints on
$\Upsilon_{\star}$. It is remarkable that the uncertainty on
$\log{r_c}$ remains large (hence rendering $r_c$ known only within
an order of magnitude). Reducing the error on $\log{r_c}$ is,
however, quite difficult with the data at hand. As can be easily
checked, $r_c$ mainly determines the value of the circular
velocity in the outer region with $v_c$ being larger for smaller
$r_c$. For $\beta \simeq 0.8$ as we will adopt later,
$(r/r_c)^{\beta} \simeq 1 - 50$ for $r$ ranging from $10^{-2}$ to
$10^{2} r_c$ where most of the data are concentrated. In order to
get smaller errors on $\log{r_c}$, one should increase the number
of points (and reducing the measurement uncertainties) in the
region $r > 10^3 r_c$. For typical values of $\log{r_c} \simeq -2$
and $R_d \simeq 2 \ {\rm kpc}$, one then needs to measure the
rotation curve beyond $r \sim 5 R_d$ which is quite unrealistic at
the moment.

\subsection{Raw vs smooth data}

In the analysis of the simulated cases described above, we have up
to now used the raw data as input to the fitting procedure.
Nevertheless, in Sect. 4, we have claimed that the smooth rather
than the raw data will be used in the analysis of the observed LSB
galaxies rotation curves. As explained above and further discussed
in de Blok \& Bosma (2002) and references therein, smooth data are
better suited to recover constraints on a given theory since they
are less sensitive to non axisymmetric features and outliers
affected by unpredictable errors.

Any smoothing procedure may potentially introduce some correlation
among the data points because of binning the data and averaging
the measurements in a given bin. Although the procedure adopted by
de Blok \& Bosma (2002) and briefly described in Appendix B is
quite efficient in reducing these problems, a residual correlation
still remains so that it is worth exploring whether this affects
the results.

To this aim, we have fitted both the raw and the smooth data with
the $\beta$ parameter set for the reasons discussed above. As a
first qualitative test, we have checked that the results are
interchangeable, i.e. we may fit equally well the raw data with
the best fit curve from the fit to the smooth data and vice versa.
This is clearly seen in the top panels of Fig. \ref{fig: simtwo}
where the long dashed lines (representing the best fit to the
smooth data) are hardly distinguished by the short dashed ones
(from the fit to the raw data). As a more rigorous test, we have
compared the best fit values from the two fits with those used to
generate the simulated curve. Indeed, we find a remarkable
agreement between the three sets of $(\log{r_c}, f_g)$ values.
What is more interesting, in many cases, $(\Delta p/p)_{smooth} <
(\Delta p/p)_{raw}$ thus advocating in favor of the use of the
smooth rather than the raw data.

As a final test, we also explore whether the confidence ranges and
hence the uncertainties on the model parameters are affected. A
nice visual result may be gained looking at the bottom panels in
Fig. \ref{fig: simtwo} and comparing the solid with the short
dashed lines. As it is clear, the confidence regions quite well
overlap with no visible offset from one another. Actually, using
smooth rather than raw data leads to wider confidence regions and
hence larger errors on $(\log{r_c}, f_g)$. However, this is
expected since the smooth dataset contains a lower number of
points so that we can roughly expect that the error $\sigma_p$ on
a parameter $p$ increases by a factor $\varepsilon \propto
(N_{raw}/N_{smooth})^{1/2}$. This is indeed the case when
comparing the estimated errors from projecting the 1D confidence
regions on the $(\log{r_c}, f_g)$ axes.

A final comment is in order. Because of how the measurement
uncertainties have been computed, the best fit reduced
$\chi^2/d.o.f$ values are not expected to be close to $1$. This is
indeed the case when dealing with the raw data. However, a further
reduction is expected for the smooth data because of the
peculiarities of the smoothing procedure used. For instance, we
get $(\chi^2/d.o.f.)_{raw} = 0.29$ vs $(\chi^2/d.o.f.)_{smooth} =
0.07$ for the case in the right panel of Fig. \ref{fig: simtwo}.
There are two motivations concurring to the finding of such small
reduced $\chi^2$. First, the uncertainties have been
conservatively estimated so that the true ones may also be
significantly smaller. Should this be indeed the case,
$\chi^2/d.o.f.$ turn out to be underestimated. A second issue
comes from an intrinsic feature of the smoothing procedure. As
discussed in Appendix B, the method we employ is designed to
recover the best approximation of an underlying model by a set of
sparse data. Since the fit to the smooth data searches for the
best agreement between the model and the data, an obvious
consequence is that the best fit must be as close as possible to
data that are by their own as close as possible to the model. As
such, if the best fit model reproduces the data, the $\chi^2$ is
forced to be very small hence originating the observed very small
values of the reduced $\chi^2$. Note that both these effects are
systematics so that they work the same way over the full
parameters space. Since we are interested in $\Delta \chi^2$
rather than $\chi^2_{min}$, these systematics cancel out thus not
affecting anyway the estimate of the uncertainties on the model
parameters.

\begin{figure*}
\centering \subfigure{\includegraphics[width=5cm]{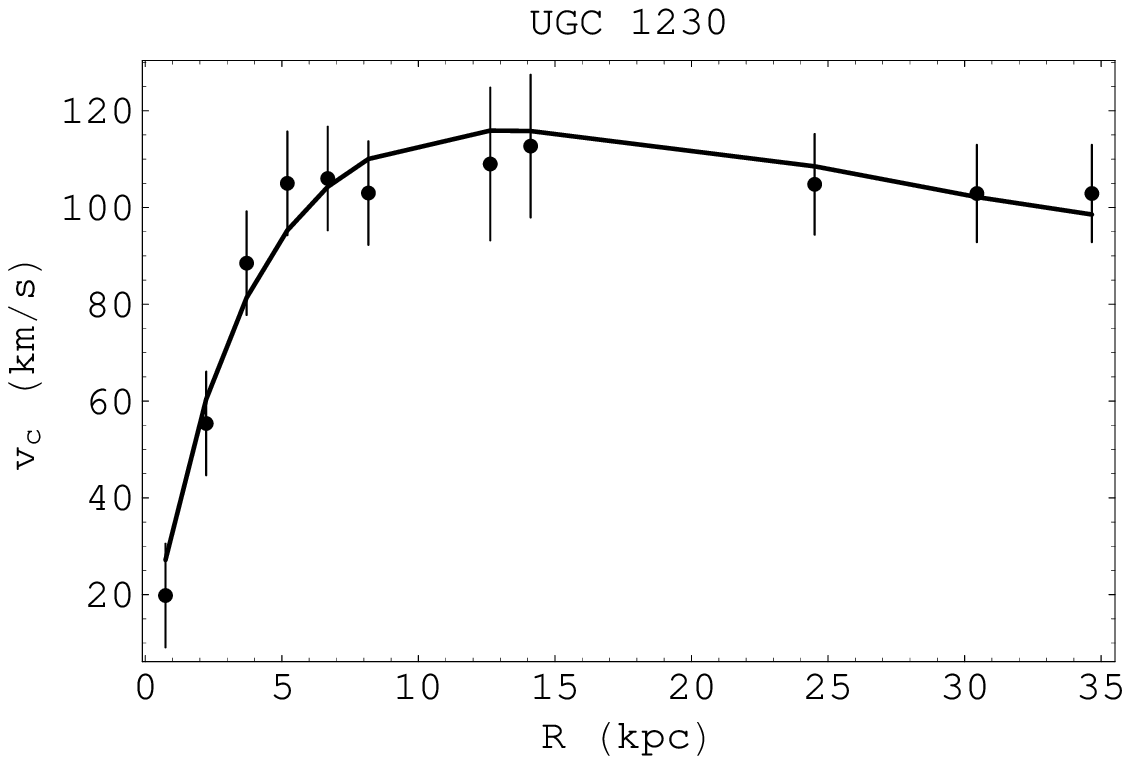}}
\goodgap \subfigure{\includegraphics[width=5cm]{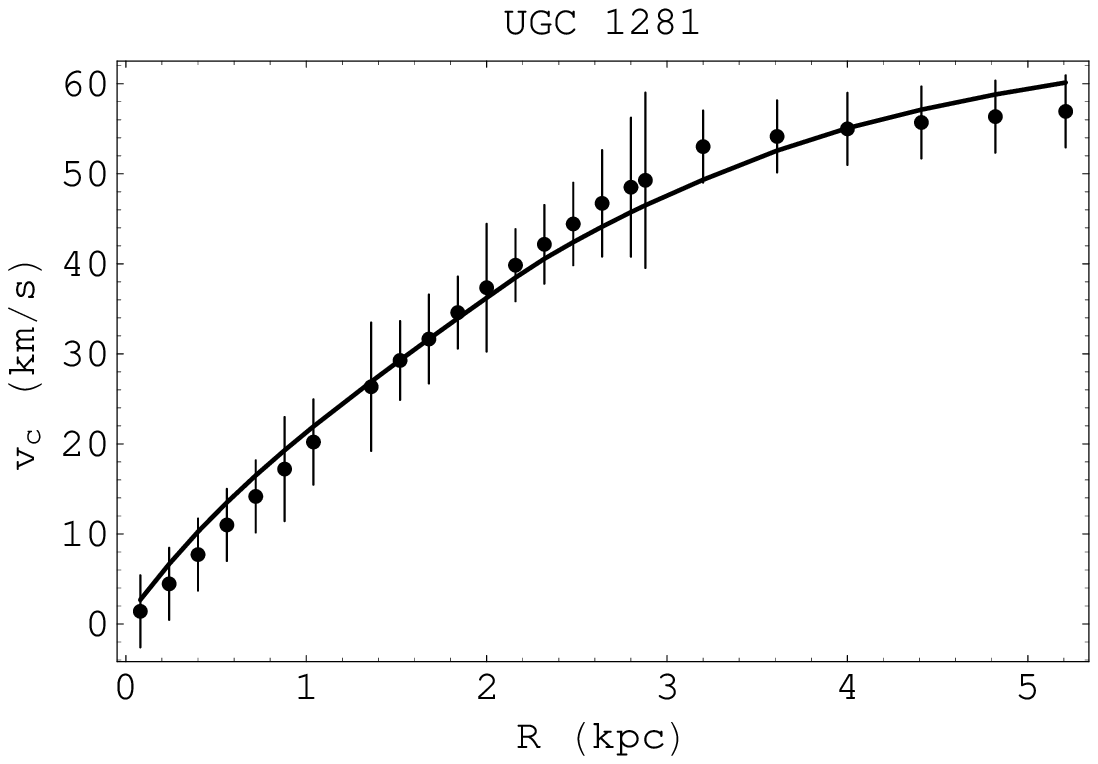}}
\goodgap
\subfigure{\includegraphics[width=5cm]{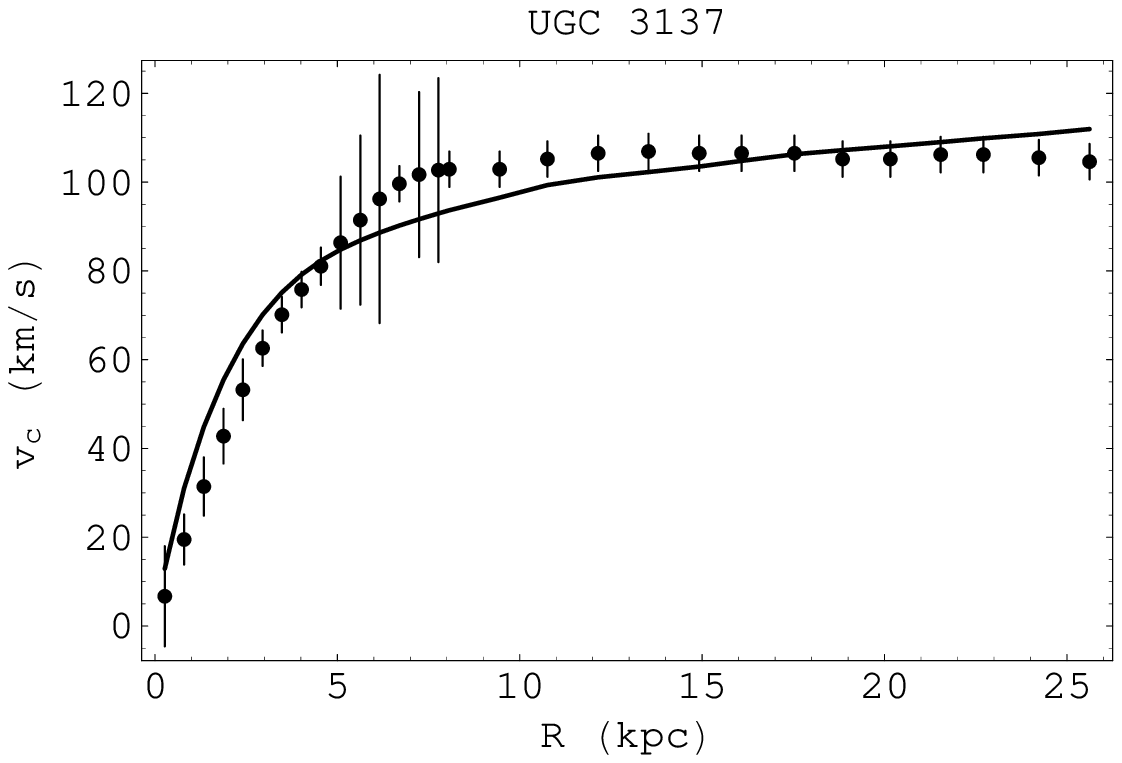}} \\
\subfigure{\includegraphics[width=5cm]{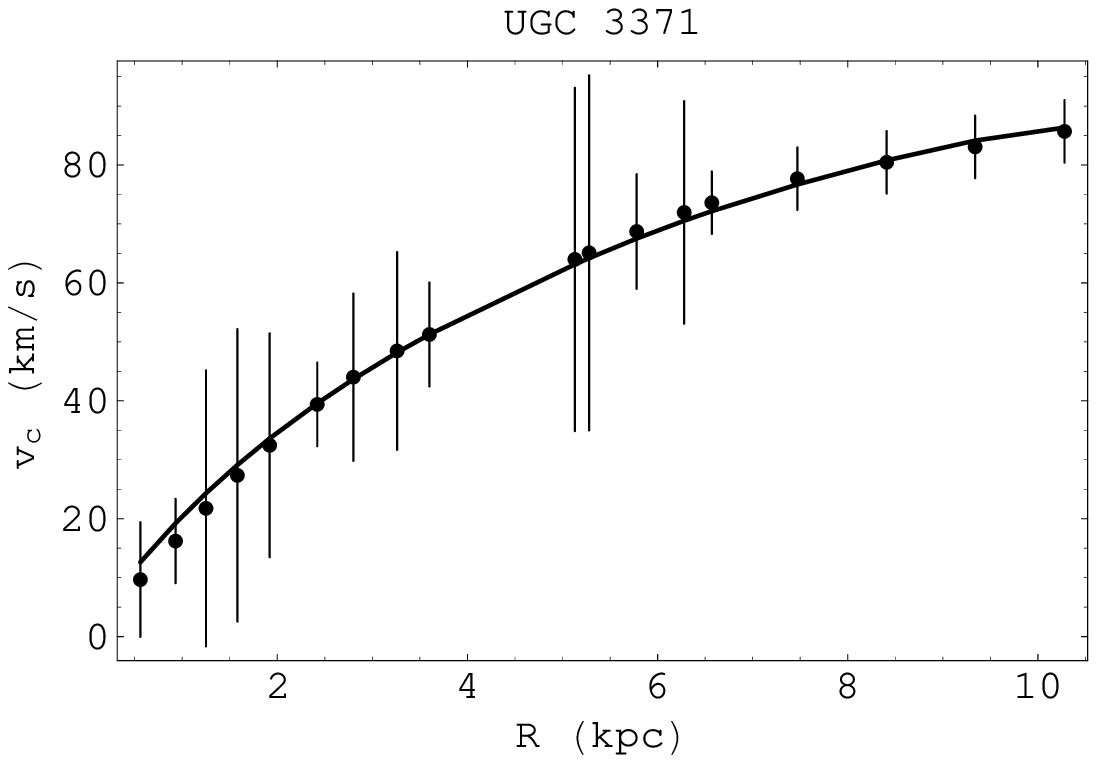}} \goodgap
\subfigure{\includegraphics[width=5cm]{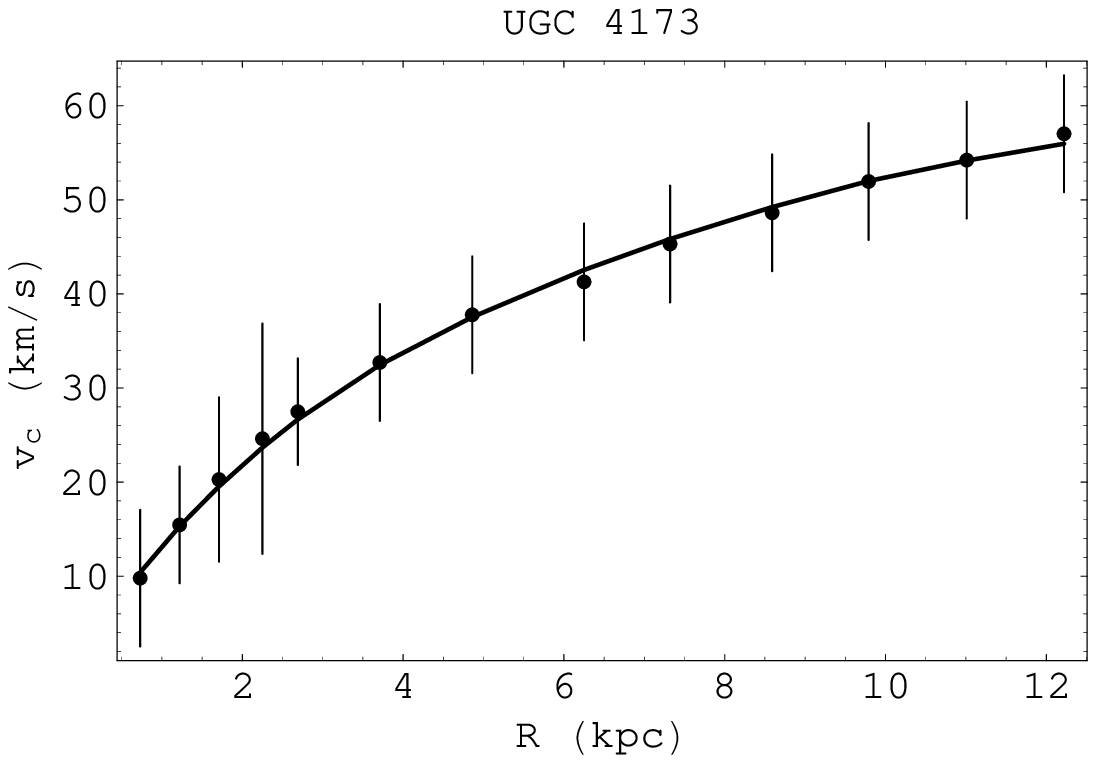}} \goodgap
\subfigure{\includegraphics[width=5cm]{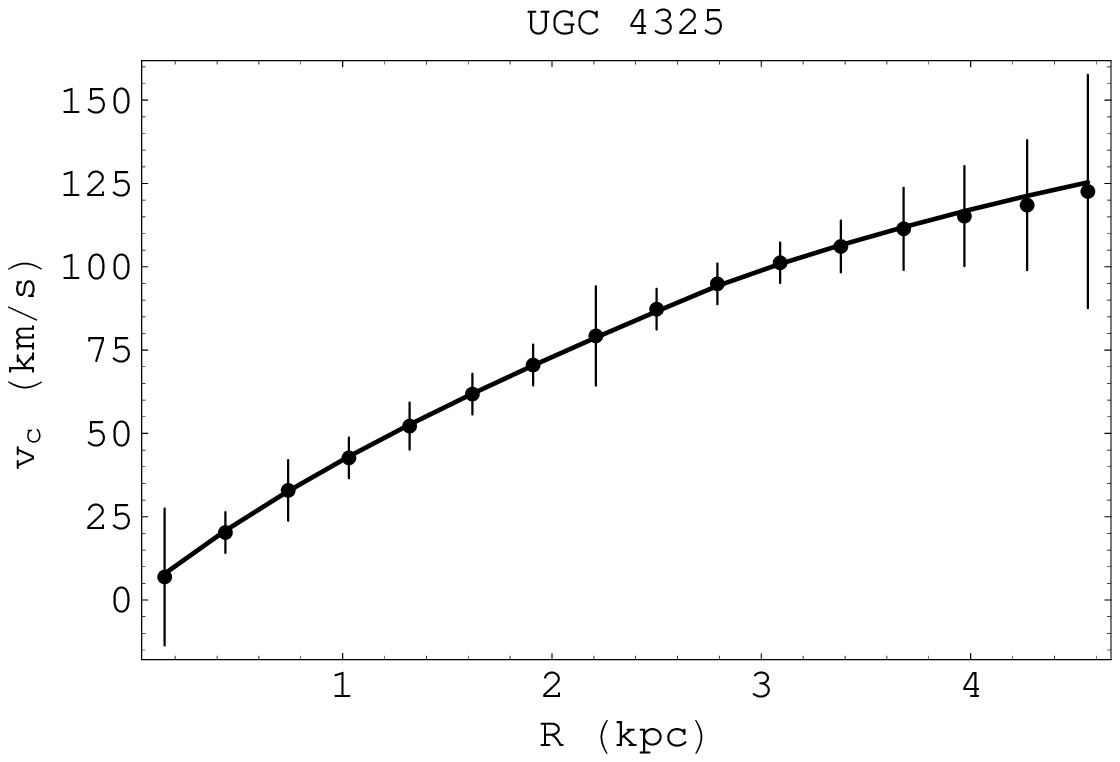}} \\
\subfigure{\includegraphics[width=5cm]{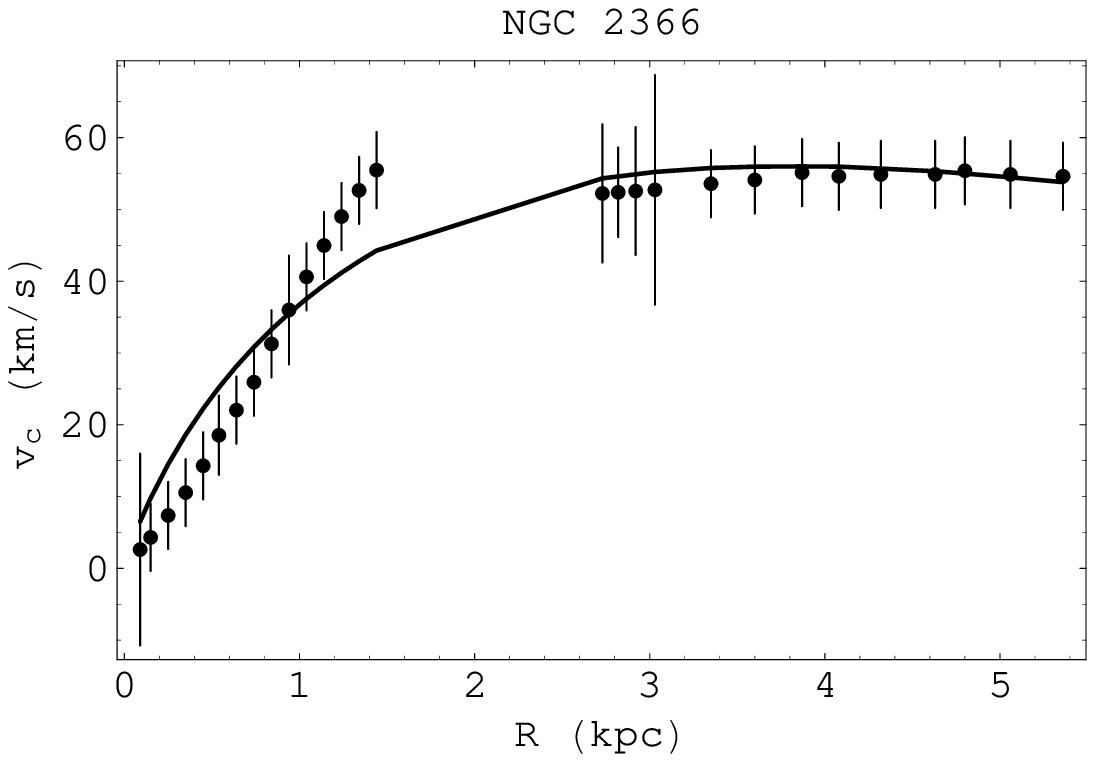}} \goodgap
\subfigure{\includegraphics[width=5cm]{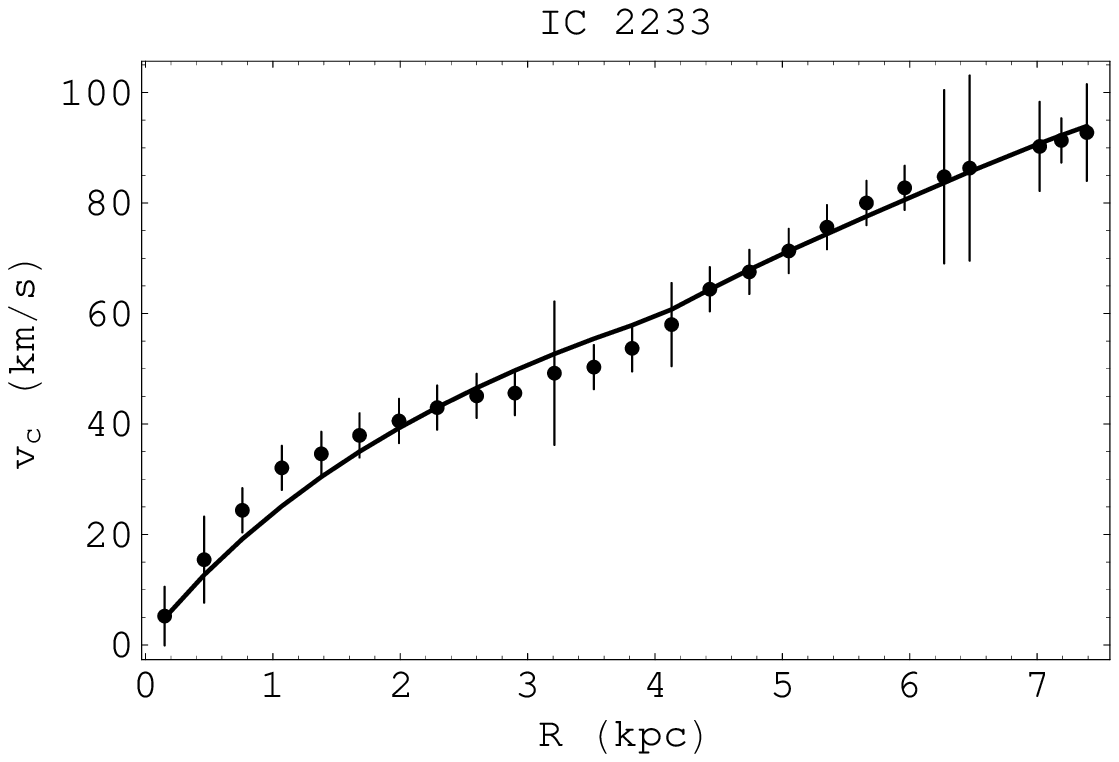}} \goodgap
\subfigure{\includegraphics[width=5cm]{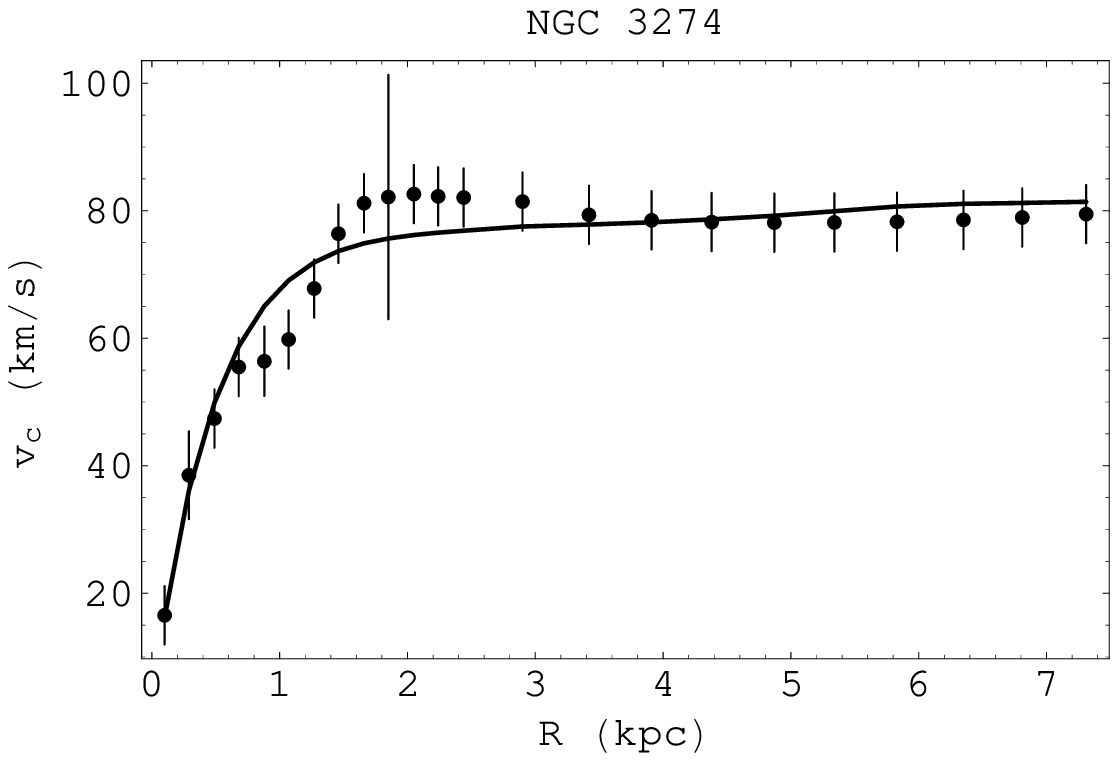}} \\
\subfigure{\includegraphics[width=5cm]{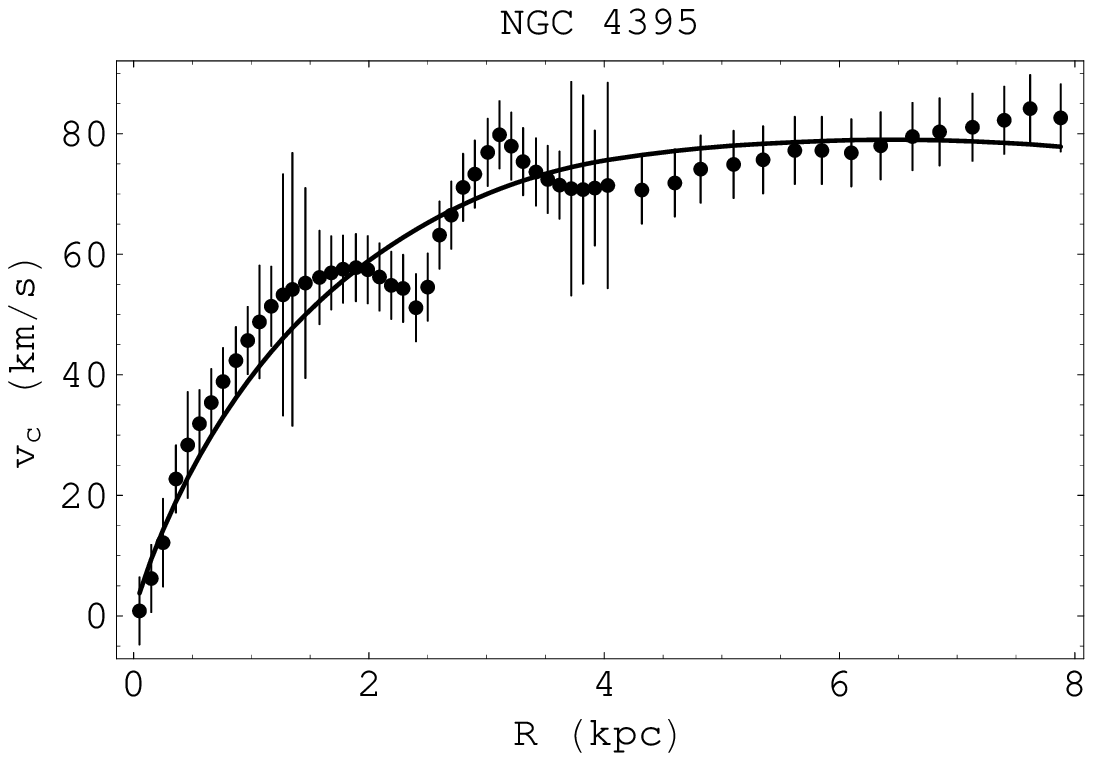}} \goodgap
\subfigure{\includegraphics[width=5cm]{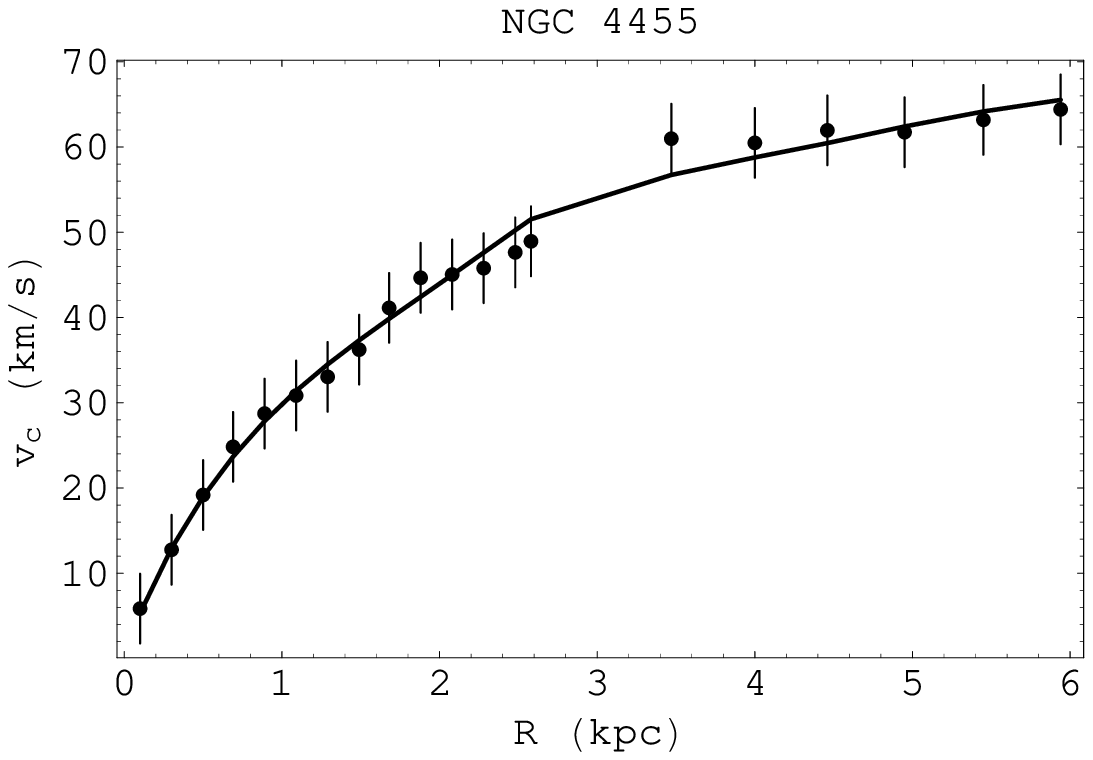}} \goodgap
\subfigure{\includegraphics[width=5cm]{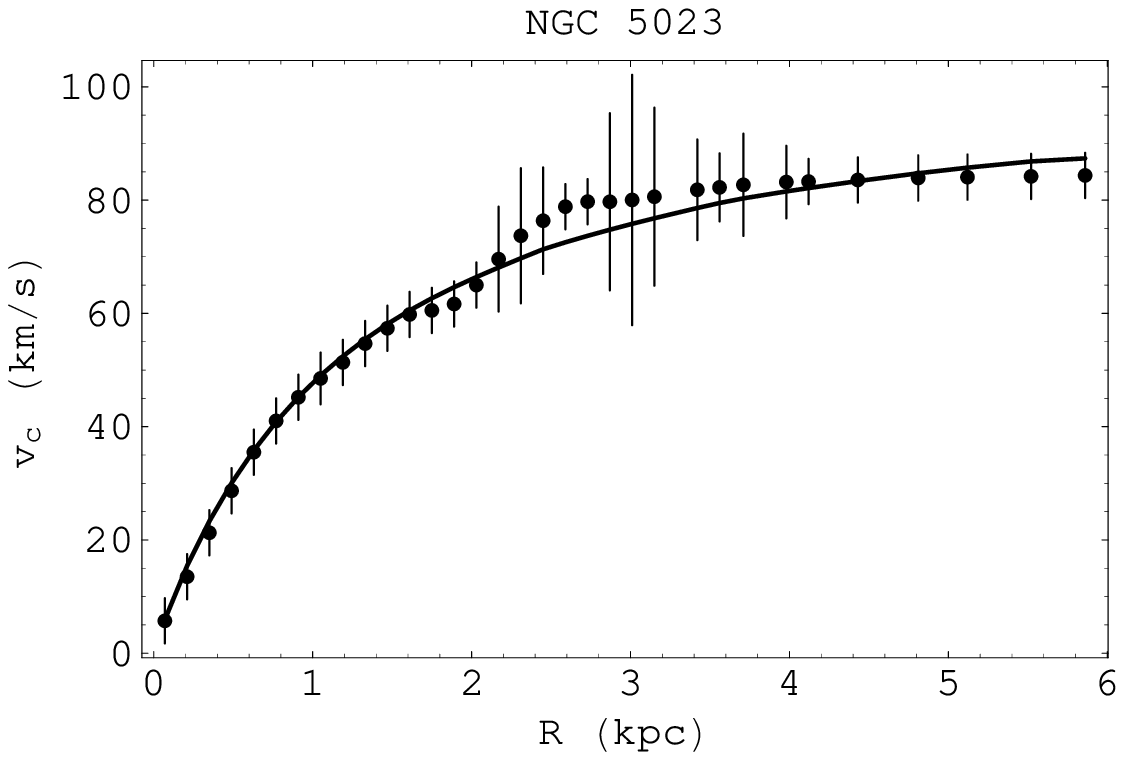}} \\
\subfigure{\includegraphics[width=5cm]{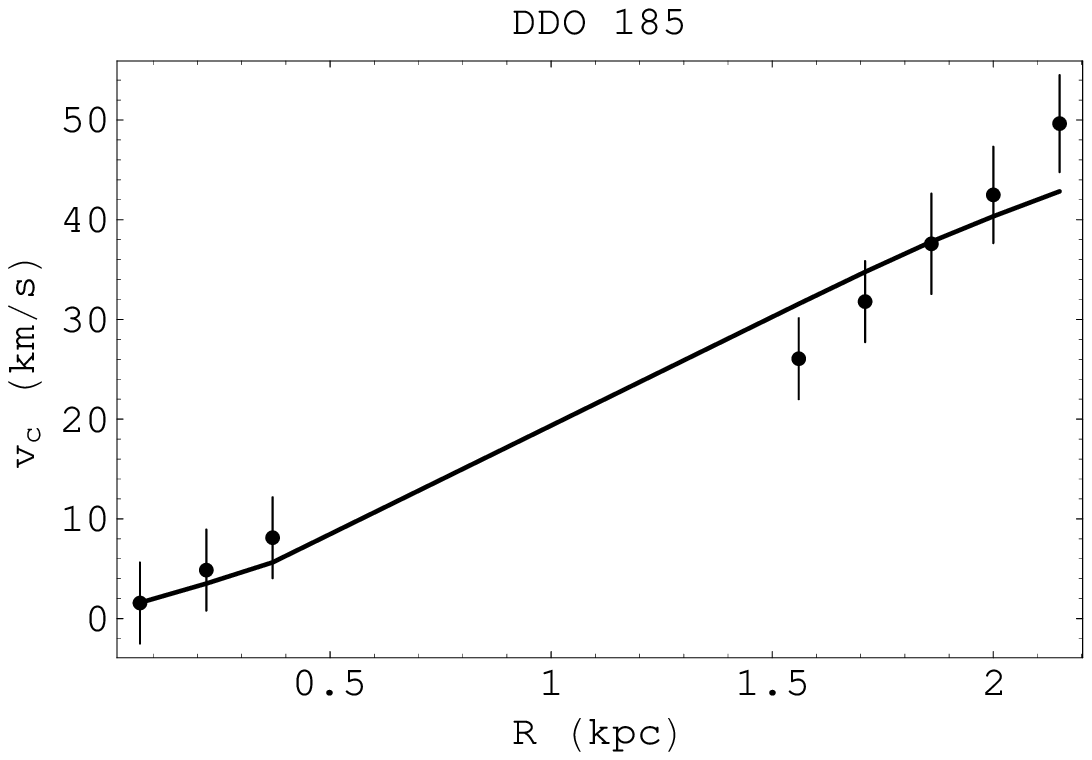}} \goodgap
\subfigure{\includegraphics[width=5cm]{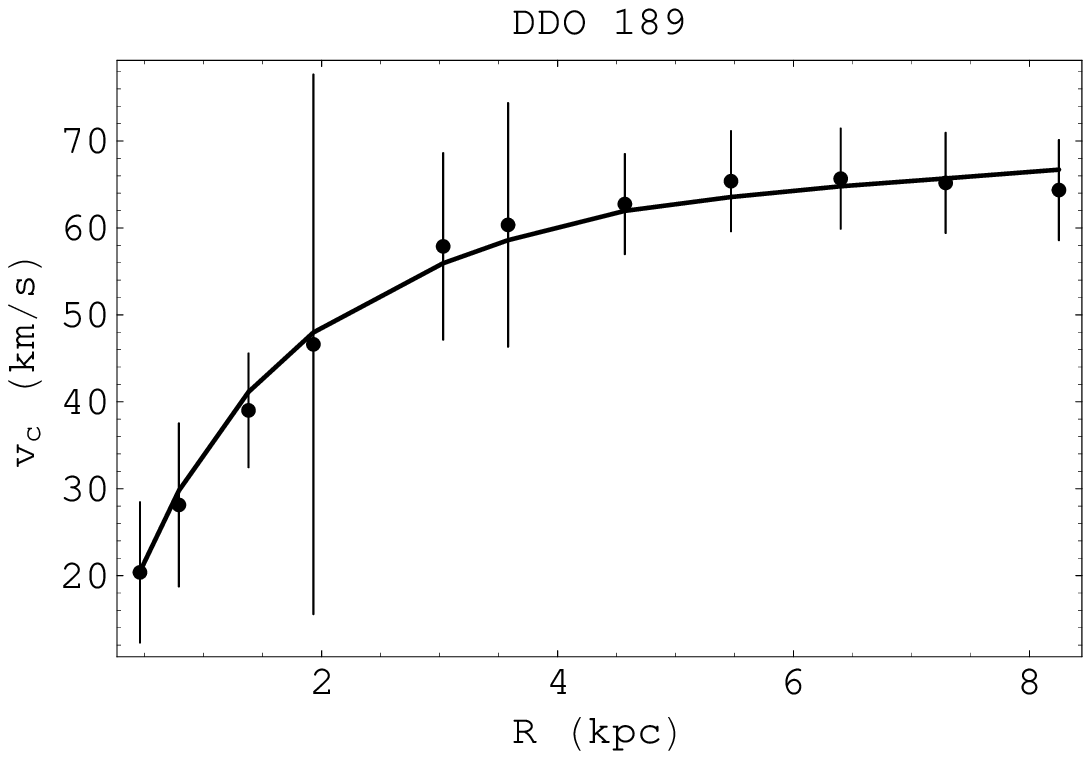}} \goodgap
\subfigure{\includegraphics[width=5cm]{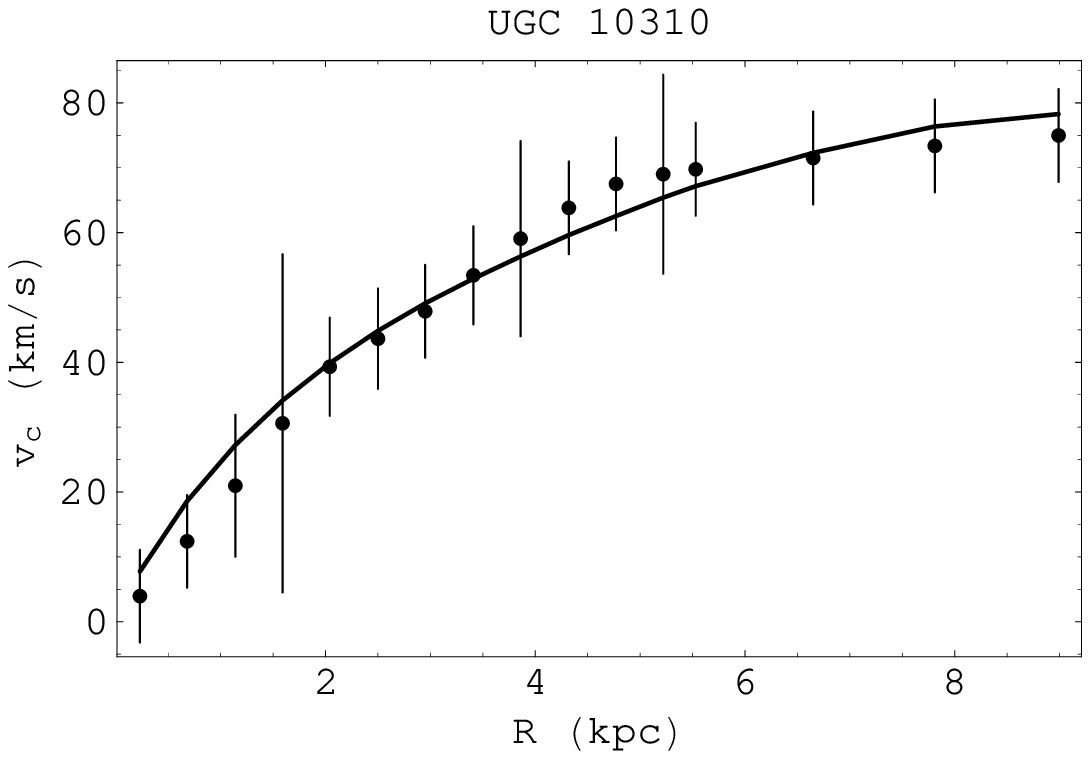}} \\
\caption{Best fit curves superimposed to the data for the sample
of 15 LSB galaxies considered. See Table 1 for details on the
galaxies and Table 2 for the values of the best fit parameters. A
case by case discussion is presented in the Appendix A.}
\label{fig: bf}
\end{figure*}

\begin{table*}
\caption{Best fit values of the model parameters from minimizing
$\chi^2(\beta, \log{r_c}, f_g)$ with $\beta = 0.817$ corresponding
to $n = 3.5$ as obtained from the best fit to the SNeIa data with
only baryonic matter. We report $1 \sigma$ $(2 \sigma)$ confidence
ranges on the fitting parameters computed by projecting on the
$(\log{r_c}, f_g)$ axes the contours $\Delta \chi^2 = 1$ $(\Delta
\chi^2 = 4)$. The best fit stellar $M/L$ ratio $\Upsilon_{\star}$
has been obtained evaluating Eq.(\ref{eq: fgdef}) for the best fit
$f_g$, while the uncertainty is obtained by usual propagation of
errors symmetrizing the $1 \sigma$ range of $f_g$. Note that this
procedure is not completely correct since the errors are not
Gaussian so that they are likely to be overestimated (especially
when giving rise to unphysical negative lower limits for
$\Upsilon_{\star}$). We also give $\chi^2/d.o.f.$ for the best fit
model.}
\begin{center}
\begin{tabular}{|c|c|c|c|c|c|c|c|c|}
\hline Id & \multicolumn{3}{|c|}{$\log{r_c}$} &
\multicolumn{3}{|c|}{$f_g$} & $\Upsilon_{\star}$ & $\chi^2/dof$ \\
\hline ~ & $bf$ & $1 \sigma$ & $2 \sigma$ & $bf$ & $1 \sigma$ & $2
\sigma$ & ~ & ~ \\ \hline \hline UGC 1230 & $-0.38$ & $(-0.59,
-0.13)$ & $(-0.78, -0.05)$ & $0.15$ & $(0.13, 0.18)$ & $(0.10,
0.21)$ & $15.9 \ {\pm} \ 3.1 \ {\pm} \ 7.1$ & 0.33 \\ UGC 1281 &
$-2.12$ & $(-2.26, -1.95)$ & $(-2.38, -1.76)$ & $0.47$ & $(0.37,
0.56)$ & $(0.29, 0.66)$ & $1.36 \ {\pm} \ 0.53 \ {\pm} \ 1.04$ &
0.22 \\ UGC 3137 & $-1.67$ & $(-1.70, -1.63)$ & $(-1.73, -0.60)$ &
$0.61$ & $(0.59, 0.63)$ & $(0.57, 0.64)$ & $12.0 \ {\pm} \ 0.9 \
{\pm} \ 1.8$ & 1.80 \\ UGC 3371 & $-1.78$ & $(-1.99, -1.52)$ &
$(-2.16, -1.21)$ & $0.40$ & $(0.28, 0.54)$ & $(0.20, 0.67)$ & $3.5
\ {\pm} \ 1.9 \ {\pm} \ 3.8$ & 0.03 \\ UGC 4173 & $-0.74$ &
$(-1.11, -0.16)$ & $(-1.39, 0.55)$ & $0.36$ & $(0.26, 0.49)$ &
$(0.20, 0.65)$ & $8.9 \ {\pm} \ 5.1 \ {\pm} \ 11.3$ & 0.01 \\ UGC
4325 & $-2.81$ & $(-2.96, -2.62)$ & $(-3.07, -2.36)$ & $0.69$ &
$(0.55, 0.80)$ & $(0.40, 0.89)$ & $0.51 \ {\pm} \ 0.33 \ {\pm} \
0.69$ & 0.01 \\ NGC 2366 & $0.03$ & $(-0.47, 1.05)$ & $(-0.77,
1.25)$ & $0.18$ & $(0.17, 0.20)$ & $(0.15, 0.23)$ & $14.4 \ {\pm}
\ 1.9 \ {\pm} \ 4.4$ & 1.09 \\ IC 2233 & $-2.05$ & $(-2.12,
-1.96)$ & $(-2.19, -1.87)$ & $0.60$ & $(0.55, 0.64)$ & $(0.50,
0.68)$ & $1.56 \ {\pm} \ 0.29 \ {\pm} \ 0.60$ & 0.50 \\ NGC 3274 &
$-2.09$ & $(-2.14, -2.03)$ & $(-2.19, -1.98)$ & $0.49$ & $(0.47,
0.52)$ & $(0.44, 0.55)$ & $2.89 \ {\pm} \ 0.30 \ {\pm} \ 0.60$ &
0.84 \\ NGC 4395 & $-0.25$ & $(-0.50, -0.05)$ & $(-0.69, 0.23)$ &
$0.093$ & $(0.090, 0.101)$ & $(0.088, 0.110)$ & $12.1 \ {\pm} \
1.6 \ {\pm} \ 2.5$ & 0.70 \\ NGC 4455 & $-2.36$ & $(-2.41, -2.30)$
& $(-2.46, -2.24)$ & $0.85$ & $(0.82, 0.87)$ & $(0.79, 0.89)$ &
$0.38 \ {\pm} \ 0.08 \ {\pm} \ 0.17$ & 0.18 \\ NGC 5023 & $-2.52$
& $(-2.58, -2.46)$ & $(-2.63, -2.40)$ & $0.52$ & $(0.49, 0.55)$ &
$(0.46, 0.58)$ & $1.02 \ {\pm} \ 0.12 \ {\pm} \ 0.26$ & 0.29 \\
DDO 185 & $-2.74$ & $(-2.81, -2.52)$ & $(-2.87, -2.10)$ & $0.94$ &
$(0.71, 0.97)$ & $(0.41, 1.00)$ & $0.12 \ {\pm} \ 0.49 \ {\pm} \
1.14$ & 0.83 \\ DDO 189 & $-1.82$ & $(-1.85, -1.47)$ & $(-2.00,
-1.24)$ & $0.53$ & $(0.43, 0.62)$ & $(0.35, 0.72)$ & $6.44 \ {\pm}
\ 2.52 \ {\pm} \ 4.90$ & 0.06 \\ UGC 10310 & $-1.76$ & $(-1.92,
-1.56)$ & $(-2.05, -1.34)$ & $0.56$ & $(0.46, 0.65)$ & $(0.37,
0.74)$ & $1.55 \ {\pm} \ 0.60 \ {\pm} \ 1.16$ & 0.21 \\
\hline
\end{tabular}
\end{center}
\end{table*}

\section{Results}

The extensive analysis of the previous section make it possible to
draw two summarizing conclusions. First, we have to set somewhat
the slope $n$ of the gravity Lagrangian in order to break the
degeneracy the model parameters. Second, we can rely on the
smoothed data without introducing any bias in the estimated
parameters or on their uncertainties.

A key role is then played by how we set $n$ and hence $\beta$. To
this aim, one may resort to cosmology. Indeed, $R^n$ gravity has
been introduced as a possible way to explain the observed cosmic
speed up without the need of any dark energy component. Motivated
by the first encouraging results, we have fitted the SNeIa Hubble
diagram with a model comprising only baryonic matter, but
regulated by modified Friedmann equations derived from the $R^n$
gravity Lagrangian. Indeed, we find that the data are consistent
with the hypothesis of no dark energy and dark matter provided $n
\ne 1$ is assumed \cite{prl}. Unfortunately, the constraints on
$n$ are quite weak so that we have decided to set $n$ to its best
fit value without considering the large error. This gives $\beta =
0.817$ that we will use throughout the rest of the paper. Note
that Eq.(\ref{eq: bnfinal}) quickly saturates as function of $n$
so that, even if $n$ is weakly constrained, $\beta$ turns out to
be less affected.

A comment is in order here. Setting $\beta$ to the value derived
from data probing cosmological scales, we are implicitly assuming
that the slope $n$ of the gravity Lagrangian is the same on all
scales. From a theoretical point of view, this is an obvious
consistency assumption. However, it should be nicer to derive this
result from the analysis of the LSB rotation curves since they
probe a different scale. Unfortunately, the parameters degeneracy
discussed above prevents us to efficiently perform this quite
interesting test. Indeed, an accurate estimate of $n$ from $\beta$
needs a well determined $\beta$ since a small offset $\Delta
\beta/\beta$ translates in a dramatically large $\Delta n/n$. As a
consequence, a possible inconsistency among the estimated $\beta$
from different galaxies could erroneously lead to the conclusion
that the gravity theory is theoretically not self consistent. To
validate such a conclusion, however, one should reduce $\Delta
\beta/\beta$ to less than $5\%$. Unfortunately, our analysis of
the simulated rotation curves have shown us that this is not
possible with the data at hand. It is therefore wiser to opt for a
more conservative strategy and look for a consistency\footnote{A
similar problem also arises when dealing with MOND where the
critical acceleration $a_0$ plays a similar role as $n$ for our
theory. In principle, one should leave this quantity free when
fitting galactic rotation curves and then check whether the same
value is recovered for all galaxies. Unfortunately, model
degeneracies prevent to perform such a test so that it is common
to set $a_0$ to its fiducial value $1.2 {\times} 10^{-10} {\rm
m/s^2}$ from the beginning.} between the results from the
cosmological and the galactic scales exploring whether the value
of $\beta$ set above allows to fit all the rotation curves with
physical values of the remaining two parameters $(\log{r_c},
f_g)$. This is our aim in this paper, while the more ambitious
task hinted above will need for a different dataset.

With this caveat in mind, all we need to fit the data is the
modelling of LSB galaxies described in Sect. 4.3 and the smoothed
data available in literature. The best fit curves thus obtained
are shown in Fig. \ref{fig: bf}, while the constraints on
$(\log{r_c}, f_g)$ and on the stellar $M/L$ ratio
$\Upsilon_{\star}$ are reported in Table 2. As a preliminary
remark, it is worth noting that three galaxies (namely, NGC 2366,
NGC 4395 and DDO 185) may be excluded by further discussion
because of problematic data.

Indeed, for NGC 2366 the lack of data in the intermediate region
prevents from deriving useful constraints, while the bump and the
sink in NGC 4395 clearly signals the effect of local clumps in the
gas distribution. Finally, for DDO 185, we have only 8 points
separated by a large gap so that the fit is unable to converge. We
stress that these cases are problematic whatever is the mass model
and the gravity theory adopted so that we will not consider them
anymore in the following discussion. A detailed
case\,-\,by\,-\,case analysis of the full sample is presented in
Appendix A, while here we mainly dedicate to some general lessons
we can draw from the fit results.

Fig. \ref{fig: bf} shows that, for 11 over 12 cases (the only
problematic one being UGC 3137), there is an overall very good
agreement between the data and the best fit curve thus suggesting
that our modified gravitational potential allows to fit the data
without any dark matter halo. Indeed, our model galaxies are based
only on what is directly observed (the stellar mass and the gas
content) and no exotic component is added. Needless to say, this
is not possible in the standard Newtonian theory of gravity, while
it is the presence of the additive power law term in the modified
gravitational potential that makes it possible to increase the
rotation curve in such a way to reproduce what is measured. In
order to further substantiate this result, we can compare the
constraints on the galactic parameters $f_g$ and
$\Upsilon_{\star}$ with what is expected from astrophysical
considerations.

First, we consider the gas mass fraction $f_g$. Roughly averaging
the best fit values for the 11 successfully fitted galaxies, we
get $\langle f_g \rangle \simeq 0.51$ with a standard deviation
$\sigma_g \simeq 0.18$. Both these values are typical of LSB
galaxies thus suggesting that our model galaxies are physically
reasonable. As a further check, one could question whether the
estimated values of the $M/L$ ratio $\Upsilon_{\star}$ are
reasonable. The stellar $M/L$ is usually obtained by fitting the
Newtonian rotation curve of the exponential disk to the observed
data in the inner region. However, such an estimate may be
seriously biased. On the one hand, one usually add a dark halo
contributing also to the inner rotation curve so that less disk
mass is needed and hence the $M/L$ ratio could be underestimated.
On the other hand, being $r_c$ of order $10^{-2} \ {\rm kpc}$,
using the Newtonian gravitational potential significantly
underestimates the {\it true} rotation curve for a given disk mass
so that more mass and hence an artificially higher $M/L$ is needed
if the halo is neglected. As a consequence, we cannot rely on the
estimates of $M/L$ reported in literature if they have been
obtained by studying the inner rotation curve. A possible way out
could be to use the relation between broad band colors and $M/L$
\cite{BdJ01}. Unfortunately, this relation has been obtained by
considering stellar population models that are typical of high
surface brightness galaxies that have quite different properties.
Moreover, such a relation has been calibrated by fitting the
Tully\,-\,Fisher law under the hypothesis of maximal disk and
Newtonian gravitational potential. Indeed, as a cross check, we
have used the Bell \& de Jong (2001) formulae with the colors
available in the NED database\footnote{Note that these colors are
typically in a different photometric system than that used by Bell
\& de Jong (2001). Although this introduces a systematic error, it
is unlikely that this causes a significant bias in the estimated
$M/L$. For details see the NED database ({\tt
http://nedwww.ipac.caltech.edu}).} obtaining values of
$\Upsilon_{\star}$ typically much smaller than 1. This is in
contrast with the usual claim that $M/L \simeq 1.4$ for LSB
galaxies \cite{dbb02}, while some suitably chosen population
synthesis models predict $\Upsilon_{\star}$ between 0.5 and 2
\cite{vdH00}.

Excluding the four problematic galaxies (UGC 3137, NGC 2366, NGC
4395, DDO 185), a direct comparison of the values of
$\Upsilon_{\star}$ in Table 2 with the fiducial range $(0.5, 2.0)$
shows that in 9 over 11 cases the fitted $\Upsilon_{\star}$ is
consistent within $1 \sigma$ with the fiducial range quoted above.
For UGC 1230 and DDO 189, the fitted $M/L$ is unacceptably high so
that a residual matter component seems to be needed. Should this
missing matter be indeed dark matter, our proposed scenario would
fail for these two galaxies. Deferring to Appendix A possible
solutions for each single case, we here note that our constraints
on $\Upsilon_{\star}$ comes from those on $f_g$ through
Eq.(\ref{eq: fgdef}). Here, an assumption on the helium fraction
$f_{He}$ has been assumed to convert the measured HI mass $M_{HI}$
into the total gas mass $M_g = f_{He} M_{HI}$. Although
reasonable, our choice for the constant conversion factor is
affected by an unknown uncertainty that we have not taken into
account. Moreover, we have assumed the same $f_{He}$ for all
galaxies, while it is conceivable that star evolution related
phenomena could make $f_{He}$ mildly galaxy dependent. Should
$f_{He}$ be lower, than $\Upsilon_{\star}$ will be smaller thus
lowering the disagreement observed. Moreover, we have not included
any molecular gas in the gas budget. Although this is typically a
good assumption, it is worth noting that our modified potential
may increase the contribution to the total rotation curve of any
mass element so that it is possible that the missing matter in UGC
1230 and DDO 189 is represented by unaccounted molecular gas.
However, even excluding these two galaxies, we end up with a
conservative estimate of 10 over 12 successful fits with plausible
astrophysical values of the fitted galactic parameters which is a
satisfactory considering the paucity of the sample.

Finally, let us consider the results on $\log{r_c}$. Different
from the case of $\beta$, $r_c$ is not a universal constant.
Nevertheless, considering the conservative sample of 9
successfully fitted galaxies (thus excluding UGC 1230, UGC 3137,
NGC 2366, NGC 4395, DDO 185, DDO 189) and roughly averaging the
best fit values, we get $\langle \log{r_c} \rangle = -2.0 {\pm}
0.6$. The reasonably low scatter in $\log{r_c}$ may be
qualitatively explained considering that $r_c$ mainly determines
the value of the terminal velocity in the rotation curve. Since
this quantity has a low scatter for the sample of LSB galaxies we
have used, it is expected that the same holds for $\log{r_c}$.

The constraints on $(\log{r_c}, f_g)$ summarized in Table 2 have
been obtained for $\beta = 0.817$, consistent with the best fit
$n$ from the fit to SNeIa Hubble diagram. However, since the
estimate of $n$ is affected by a large uncertainty so that it is
worth wondering how this impacts the results presented here. To
this aim, we have repeated the fit for UGC 10310 for $n = 2.5$
$(\beta = 0.740)$ and $n = 4.5$ $(\beta = 0.858)$. For the best
fit values, we get\,:

\begin{displaymath}
(\log{r_c}, f_g, \Upsilon_\star) = (-1.85, 0.58, 1.42) \ \ for \
\beta = 0.740 \ ,
\end{displaymath}

\begin{displaymath}
(\log{r_c}, f_g, \Upsilon_\star) = (-1.36, 0.41, 1.62) \ \ for \
\beta = 0.858 \ ,
\end{displaymath}
to be compared with $(\log{r_c}, f_g, \Upsilon_\star) = (-1.76,
0.56, 1.55)$. As expected, increasing $\beta$, $\log{r_c}$ and
$f_g$ become smaller in order to give the same observed rotation
curve, consistent with what expected from Fig.\,\ref{fig: vcont}.
Although the shift in the best fit values is significant, the data
do not still allow to draw a definitive conclusion. For instance,
the $2 \sigma$ confidence ranges for $\log{r_c}$ turn out to be\,:

\begin{equation}
\begin{array}{ll}
(-2.16, -1.39) & for \ \beta = 0.740 \\
(-2.05, -1.34) & for \ \beta = 0.817 \\
(-2.04, -1.35) & for \ \beta = 0.858 \\
\end{array}
\nonumber
\end{equation}
which are consistent with each other. Note that further increasing
$n$ have no significant effect on the estimate of the parameters
since $\beta$ quickly saturates towards its asymptotic value
$\beta = 1$. We are therefore confident that, although the
constraints on $(\log{r_c}, \beta)$ depend on $\beta$, our main
results are qualitatively unaltered by the choice of $n$ (and
hence $\beta$).

It is also worth noting that the three values of $\beta$
considered above all provide quite good fits to the observed
rotation curve. This is not surprising given the data at hand and
our analysis parameters degeneracies presented in Sect. 5. In
order to constrain $\beta$ from rotation curves leaving it as a
free parameter in the fit, therefore, one could explore the
possibility to performed a combined $\chi^2$ analysis of the full
set of rotation curves. This can eventually be complemented by
adding a prior on $\beta$, e.g., from the cosmological constraints
on $n$. Exploring this issue is outside our aims here, but should
be addressed in a forthcoming paper.

Summarizing, the results from the fit and the reasonable agreement
between the recovered $\Upsilon_{\star}$ and that predicted from
stellar population synthesis models make us confident that $R^n$
gravity is indeed a possible way to fit the rotation curves of LSB
galaxies using only baryonic components (namely, the stellar disk
and the interstellar gas) thus escaping the need of any putative
dark matter halo.

\section{Discussion and Conclusions}

Rotation curves of spiral galaxies have been considered for a long
time the strongest evidence of the existence of dark matter haloes
surrounding their luminous components. Notwithstanding the great
experimental efforts, up to now there has never been any firm
detection of such an exotic dark component that should make up
these haloes. It is therefore worth wondering whether dark matter
indeed exists or it is actually the signal of the need for a
different gravitational physics.

Motivated by these considerations, we have explored here the case
of $R^n$ gravity. Since such theories have been demonstrated to be
viable alternatives to the dark energy giving rise to scenarios
capable of explaining the observed cosmic speed up, it is highly
interesting to investigate their consequences also at galactic
scales. To this aim, we have solved the vacuum field equations for
power\,-\,law $f(R)$ theories in the low energy limit thus
deriving the gravitational potential of a pointlike source. It
turns out that both the potential and the rotation curve are
corrected by an additive term scaling as $(r/r_c)^{\beta - 1}$
with the scalelength $r_c$ depending on the system physical
features (e.g. the mass) and $\beta$ a function of the slope $n$
of the gravity Lagrangian. In particular, for $n = 1$, our
approximated solution reduces to the standard Newtonian one. For
$0 < \beta < 1$, the potential is still asymptotically vanishing,
but the rotation curve is higher than the Newtonian one. These
results still hold when we compute the corrected potential for
extended systems with spherical symmetry or thin disk
configuration. As a result, we have argued that the rotation curve
of spiral galaxies could be fitted using the luminous components
only thus eliminating the need for dark matter.

In order to verify this hypothesis, we have considered a sample of
15 LSB galaxies with well measured combined HI and H$\alpha$
rotation curves extending far beyond the optical edge of the disk.
Since these systems are usually claimed to be dark matter
dominated, reproducing their rotation curves without the need of
any dark matter halo would represent a significant evidence in
favour of $R^n$ gravity. Moreover, fitting to rotation curves
allows in principle to constrain the theory parameters $(\beta,
r_c)$ and determine the $M/L$ ratio of the stellar component.
Unfortunately, extensive simulations have highlighted the need to
set a strong prior on $\beta$ (and hence $n$) to break the
degeneracy among the three fitting parameters $(\beta, \log{r_c},
f_g)$. To this aim, we have resorted to the results of SNeIAa
Hubble diagram fitting without dark matter and dark energy which
shows that $n = 3.5$ reproduces the data without the need of any
dark sector.

Motivated by this consideration, we have set $n = 3.5$ giving
$\beta = 0.817$ in order to check whether $R^n$ gravity may fit
both the SNeIa Hubble diagram and the LSB rotation curves without
either dark energy on cosmological scales and dark matter on
galactic scales with the same value of the slope $n$. Indeed, we
conservatively estimate that 10 of a sample of 12 usable galaxies
can be properly fitted by the corrected rotation curves based only
on the baryonic components (stars and gas) of the galaxies with
values of the $M/L$ ratio which may be reconciled with predictions
from stellar population synthesis models. It is worth emphasizing
that all the LSB rotation curves have been successfully fitted
using the same value of $\beta$. Although $\beta$ has been set
from the beginning, this does not guarantee that the full set of
curves will be satisfactorily well fitted. Indeed, should we have
found that a single rotation curve demands for a different
$\beta$, this could have been a decisive evidence against $R^n$
gravity. On the contrary, the same $\beta$ leads to equally good
fit for all the 10 successful galaxies. We therefore conclude that
the self consistency of the theory has been verified thus leading
further support to $R^n$ gravity as a viable alternative to the
dark sector on galactic and cosmological scales.

These encouraging results are a strong motivation for
investigating $R^n$ gravity further from both observational and
theoretical point of views. Still remaining on galactic scales, it
is mandatory to extend the analysis of the rotation curves to the
case of high surface brightness (HSB) galaxies. Although their
structure is more complicated (since one has to include also a
bulge component), HSB galaxies are more numerous than LSB ones so
that we may perform our test on a larger sample thus increasing
the significance of the results. To this aim, it is important to
carefully select the sample in order to include systems with well
measured and extended rotation curve and not affected by possible
non circular motions due to spiral arms or bar\,-\,like
structures. While this could be a limitation, it is worth
stressing that in modeling HSB one may neglect the gas component
which has been the most important source of theoretical
uncertainty in our study of LSB galaxies. Should the test on HSB
be successful as the present one, we could convincingly
demonstrate that $R^n$ gravity is a {\it no dark matter} solution
to the long standing problem of the rotation curves of spiral
galaxies.

As well known, dark matter is invoked also on larger than galactic
scales. For instance, dark matter haloes are typically present in
clusters of galaxies and enter in a crucial way in determining the
gas temperature profile which is measured from the X\,-\,ray
emission. In such a case, the form of the gravitational potential
plays a key role so that it is worth investigating whether our
modified potential could reproduce the observed temperature
profile without the need of dark matter. This test should also
represent a further check of the consistency of the theory since
it allows a determination of $\beta$ on a completely different
scale. Of course, one should find the same $\beta$, while a
significant difference could be a clear signal of unescapable
problems. Note that there are nowadays a large number of clusters
whose gas temperature profiles is well measured thanks to the {\it
Chandra} and {\it XMM\,-\,Newton} satellites so that also this
test could be performed on a large sample to ameliorate the
statistics.

A step further leads us to the cosmological scales where dark
matter is introduced to fill the gap between the baryonic density
parameter $\Omega_b$ and the estimated total matter one
$\Omega_M$. According to nucleosynthesis predictions, $\Omega_b =
0.0214 {\pm} 0.0020$ \cite{kirk}, while $\Omega_M \simeq 0.25$ is
estimated by SNeIa Hubble diagram and matter power spectrum. Such
a large discrepancy may seem to be impossible to cure without
resorting to dark matter. However, this could also not be the case
when considering that both the SNeIa Hubble diagram and the matter
power spectrum are usually computed assuming General Relativity as
the correct theory of gravity. However, should $f(R)$ theories be
indeed the correct model, one should recompute {\it ab initio} the
matter power spectrum so that it is impossible to predict {\it a
priori} what is the value of $\Omega_M$ that allows a nice fit to
the measured matter power spectrum in a higher order theory. There
is therefore much room for further investigation and it is indeed
possible that a baryons only universe is in agreement with the
cosmological data if an alternative theory of gravity, instead of
Einstein General Relativity, is used.

As a final comment, we would like to stress the power of an
approach based on higher order theories of gravity. Although it is
still possible that the choice $f(R) = f_0 R^n$ is unable to
positively pass all the tests we have quoted above, it is
important to note that $f(R)$ theories are the unique mechanism
able to explain in a single theoretical framework physical
phenomena taking place on widely different scales. In our opinion,
therefore, if a unified solution of the dark matter and dark
energy problems exist, this is the realm where it has to be
searched for with the greatest chances of successful results. \\

{\it Acknowledgements.} We warmly thank M. Capaccioli for the
interesting discussions and encouragements. We are also extremely
grateful to W.J.G. de Blok for his prompt answers to the many
questions on the data and the LSB modeling, to G. D'Agostini for
an illuminating discussion on the use of marginalized likelihood
functions and to F. Giubileo for help with data retrieving. We
also thank  G. Lambiase, C. Rubano and G. Scarpetta for
discussions and comments on preliminary versions of the paper.
Finally, the referee is greatly acknowledged for his suggestions
that have helped to radically improve the paper.

\appendix

\section{Details on fit results}

In Sect. 6, we have discussed the main features of the fit results
as a whole, while here we give some few details on the comparison
of the model with the rotation curve on a case\,-\,by\,-\,case basis. \\

{\it UGC 1230.} This is a somewhat problematic case giving a best
fit $\Upsilon_{\star} = 15.9 {\pm} 3.1$ which is hard to explain
in terms of reasonable population synthesis models. High values of
$M/L$ are also obtained in the case of maximum disk and dark halo
models. For instance, de Blok \& Bosma (2002) find
$\Upsilon_{\star} = 6.1$ for both isothermal and NFW dark halo
models. It is therefore likely that a problem may reside in the
data or in the modelling (e.g., an underestimate of the gas
content or a wrong measurement of the distance of the galaxy that
could lead to underestimate the total absolute luminosity and
hence overestimating $\Upsilon_{\star}$). Although such a
possibility exist, it is worth noting that the value of
$\log{r_c}$ is significantly larger than what is found for the
other galaxies thus enhancing the need for an unseen component at
odds with our working hypothesis of no dark matter. We have
therefore decided to not consider UGC 1230 as a successful fits
even if there is a good agreement between the data and the best fit curve. \\

{\it UGC 1281.} This is a typical case with the model nicely
reproducing the data and a value of $\Upsilon_{\star}$ in
agreement with population synthesis models. There is only a
marginal overestimate of the rotation curve for $R \le 1 \ {\rm
kpc}$, but it is well within the errors. To this aim, we remark
that a slight overestimate of the theoretical rotation curve for
the innermost points is expected for all galaxies since we have
artificially assumed the gas surface density is flat in this
region where no data are available. Should this not be the actual
situation, $v_c$ turns out to be slightly biased high. \\

{\it UGC 3137.} This case is not satisfactory for our approach.
Indeed, the reduced $\chi^2$ for the best fit model is anomalously
high ($\sim 2$) essentially due to the theoretical curve being
higher than the observed one for the innermost points and smaller
in the intermediate region. Moreover, the estimated
$\Upsilon_{\star} = 12.0 {\pm} 1.9$ is too large to be reconciled
with population synthesis models. The disagreement is hard to
explain given that the data seem to be of good quality and the
curve is quite smooth. It is, however, worth noting that it is not
possible to achieve a good fit also in the dark matter case
whatever is the halo model used (see, e.g., de Blok \& Bosma
2002). It has to be remarked that UGC 3137 is an edge\,-\,on
galaxy so that deriving a disk mass model from the surface
brightness involves a series of assumptions that could have
introduced some unpredictable systematic error. \\

{\it UGC 3371.} The agreement between the data and the model is
extremely good and the estimated values of $(\log{r_c}, f_g)$ are
typical of the sample we have considered. The best fit
$\Upsilon_{\star}$ is somewhat larger than expected on the basis
of stellar population model, but the fiducial $\Upsilon_{\star} =
1.4$ typically used in dark matter fitting is only $1 \sigma$
smaller. We can therefore consider this fit successfull and
physically reasonable. \\

{\it UGC 4173.} Although the agreement between the data and the
best fit model is almost perfect, this case is somewhat more
problematic than UGC 3371 since we get an anomalously high
$\Upsilon_{\star}$. As such, we could deem this galaxy as a
failure for $R^n$ gravity. However, examining the 2D confidence
regions in the plane $(\log{r_c}, f_g)$, it is easy to find out
models with typical values of $\log{r_c}$ and lower
$\Upsilon_{\star}$ that could still agree with the rotation curve
within the errors. Moreover, the uncertainties on the data points
are probably overestimated as could be inferred noting that also
dark halo models reproduce the observed curve with a very small
$\chi^2$ which is a typical signal of too high errors. Given these
issues, we include this galaxy in the sample of successful
fits. \\

{\it UGC 4325.} The best fit model matches perfectly the observed
rotation curve with a typical value of $\log{r_c}$, but a somewhat
small but still reasonable $\Upsilon_{\star}$. However, since
$\log{r_c}$ and $\Upsilon_{\star}$ are positively correlated, one
could increase $\log{r_c}$ and hence $\Upsilon_{\star}$ still
achieving a very good fit to the data, even if this is not our final choice. \\

{\it NGC 2366.} This curve is a challenge both for $R^n$ gravity
and dark matter models. The very linear rise in the inner part
rapidly changes in a flat part at larger radii. Moreover, there
are no points in the intermediate region that could give
constraints on how the change takes place. As de Blok \& Bosma
(2002) suggests, it is possible that the outermost points which
are based on the HI data alone are underestimated. Another
possibility is the presence of non circular motions due to the
inner bar\,-\,like structure. These uncertainties on the data lead
to a very bad fit with large values for both $\log{r_c}$ and
$\Upsilon_{\star}$. Given this situation, we have not considered
anymore this galaxy stressing, however, that this is not an evidence against $R^n$ gravity. \\

{\it IC 2233.} It is quite difficult to get a very good fit to
this galaxy rotation curve since, looking at the plot, one sees an
abrupt change of concavity for $R \ge 3 \ {\rm kpc}$. As a
consequence, a perfect matching between the data and the model is
not possible. Nevertheless, the best fit model provides a quite
good agreement with the data. Moreover, the best fit values of
$(\log{r_c}, f_g)$ are typical of our sample and the estimated
$\Upsilon_{\star}$ nicely agrees with the fiducial one suggested
in previous literature. We can therefore consider this galaxy as
one of the most remarkable successes of $R^n$ gravity. \\

{\it NGC 3274.} Although the general trend of the curve is well
reproduced, there is a certain disagreement in the region $1 \
{\rm kpc} \le R \le 2 \ {\rm kpc}$ where a change of concavity
occurs that is not reproduced by the model. Note that features
like this could be related to a clumpiness in the gas distribution
that cannot be modeled analytically. Considering, moreover, that
the value of $\log{r_c}$ is quite typical and the estimated $M/L$
ratio is not too difficult to reconcile with population synthesis
models (although somewhat high), we conclude that $R^n$ gravity
successfully reproduces this curve. \\

{\it NGC 4395.} The rotation curve of this galaxy is strongly
affected by the presence of star formation regions that cause an
oscillating behaviour for $1.5 \ {\rm kpc} \le R \le 4 \ {\rm
kpc}$ that is not possible to reproduce by any analytical model.
Indeed, the best fit model is unable to agree reasonably well with
the data so that the results on $(\log{r_c}, f_g)$ are
significantly altered. Given the problems with the modeling, we
have therefore decided to exclude this galaxy from the final
sample since it is impossible to decide whether a bad fit is a
signal of a breakdown for $R^n$ gravity. \\

{\it NGC 4455.} Both the fit and the estimated values of the model
parameters are quite satisfactory, although the low
$\Upsilon_{\star}$ may argue in favour of a higher $\log{r_c}$.
Note that there is a hole in the observed rotation curve around
$\sim 3 \ {\rm kpc}$. Adding some more data in this region could
help in better constraining the parameters with particular regard
to $\Upsilon_{\star}$. It is worth noting that the best fit curve
tends to be higher (but well within the measurement errors) in the
outer region. Extending the measurement of this galaxy rotation
curve to larger radii could therefore be a crucial test for our
paradigm in this particular case. Note, however, that it is also
possible that the parameters will be adjusted in such a way to still provide a good fit. \\

{\it NGC 5023.} This edge\,-\,on galaxy is, in a certain sense, an
ameliorated version of UGC 3137. Indeed, the best fit model
underestimates the rotation curve in the region between 2 and 3
kpc, but fits quite well the remaining data. Inspecting the
rotation curve, a change of concavity occurs at 2 kpc and it is,
indeed, this feature the origin of the disagreement. The
similarity with the case of UGC 3137 could suggest to reject this
galaxy considering also this fit as an unsuccessful one. However,
a closer look shows that, while in the case of UGC 3137 the best
fit model works bad both in the inner and the intermediate
regions, here the disagreement is limited to the zone where the
change of concavity takes place. Moreover, in this case, the best
fit $\log{r_c}$ is typical of our sample and the estimated $M/L$
ratio is quite reasonable so that we have finally decided to retain this galaxy. \\

{\it DDO 185.} This very linear curve is quite difficult to
reproduce and, indeed, our best fit model makes a poor job with a
too small $M/L$ ratio. However, the overall rotation curve
measurements are of very poor quality so that this galaxy can be discarded from further considerations. \\

{\it DDO 189.} There is an almost perfect matching between the
data and the best fit model. The estimated values of $(\log{r_c},
f_g)$ are typical for the LSB galaxies in our sample, but
$\Upsilon_{\star}$ is unexpectedly large. Since $f_g$ takes a
completely reasonable value, a possible problem could arise with
the conversion from $f_g$ to $\Upsilon_{\star}$. For instance,
should $f_{He}$ be smaller than our fiducial value, then
$\Upsilon_{\star}$ should be revised towards lower values. In
order to be conservative, we have however decided to exclude this
galaxy from the sample of successful fits even if nothing prevents
the reader to take the opposite decision. \\

{\it UGC 10310.} Everything works well for this galaxy. The best
fit model provides a good fit to the observed rotation curve with
only a modest overestimate (still within the uncertainties) in the
innermost region that could be ascribed to our assumptions in the
gas modelling. The values of $\log{r_c}$ and $f_g$ are typical of
our sample, while the best fit $\Upsilon_{\star}$ may be easily
reconciled with the predictions from stellar population synthesis
models.

\section{Smoothing the data}

As input to the pipeline for testing $R^n$ gravity, we have used
the smoothed version of the rotation curve data following the same
approach used by de Blok \& Bosma (2002). As explained in Sect. 4,
the use of the smooth rather than the raw data relies on the need
to eliminate any residual effect of non circular motions on the
data since the gas and disk mass models assume that these
components are axisymmetric systems. However, the smoothing
procedure may in principle introduce correlations among the data
so that it is worth investigating whether this may bias somewhat
the results on the model parameters. This task has been
extensively performed in Sect. 5.3 so that here we only give some
more details on how the smoothing is done referring to Loader
(1999) and de Blok \& Bosma (2002) for further details.

First, the data have been symmetrized and folded and then
resampled each 6 arcsec. Since both HI and H$\alpha$ data are
available, when there is an overlap, only H$\alpha$ points have
been retained for $r \le r_\alpha$ with the values of $r_\alpha$
available in Table 1 of de Blok \& Bosma (2002). The smooth data
have been obtained using a local regression method following the
steps schematically sketched below.

\begin{enumerate}

\item{Choose $N$ points ${x_1, \ldots, x_N}$.}

\item{Define the weight function $w(u) = (1 - |u|^3)^3$ for $|u| \le 1$, $w(u) = 0$ for $|u| > 1$,
with $u = (x - x_i)/h$ being $h$ the width of the bin centred on
$x_i$.}

\item{Given a set of $N_{obs}$ observed points $(x_j, v_j)$, perform a weighted fit of
the polynomial ${\cal{P}}_p = \sum_{l = 0}^{p}{a_l (x - x_i)^l}$
to these data using the function $w(u)$ defined above to generate
the weights.}

\item{Estimate the smoothed value of the rotation curve in $x_i$
as the zeroth order term of the polynomial fitted above.}

\item{Repeat this procedure for each one of the $N$ points $(x_1, \ldots, x_N)$
thus ending with a smoothed rotation curve dataset made out of $N$
points $(x_i, v_i)$.}

\end{enumerate}
It is worth stressing that this smoothing procedure introduces a
negligible amount of correlations among the different points in
the rotation curve. Indeed, the fit of the polynomial of degree
$p$ is performed only locally so that two consecutive bins have
only a modest (if not null) number of points in common depending
on the chosen value of $h$. As discussed in Loader (1999), local
regression methods make it possible to recover the information
about a given function eliminating the noise that affects the
observed data. At the same time, being local, the regression
method minimizes the correlations between the different bins.

As shown in Sect. 5.3, the best fit model passes almost perfectly
within the simulated smoothed data yielding a very small $\chi^2$
value. This is an expected consequence of the ability of local
regression methods to recover the true model from noisy data. As
such, if the fitted model reproduces the data, it is, in a certain
sense, forced to pass almost exactly through the smoothed points
since they represent the best approximation to the true underlying
model. If the fit is successful, the true model and the best fit
model coincide so that the $\chi^2$ is expected to be very small.

\end{document}